\documentclass[twocolumn,dvipsnames]{aastex63}

\usepackage{graphicx}
\usepackage{amsmath}
\usepackage{amssymb}
\usepackage{hyperref}
\usepackage{multirow}

\usepackage[utf8]{inputenc}
\usepackage{CJK}

\renewcommand{\deleted}[1]{{}}

\newcommand{\Msun}{\mathinner{{\rm M}_\odot}}
\newcommand{\Mpch}{\mathinner{{\rm Mpc}/h}}
\renewcommand{\vec}[1]{{\mathinner{\boldsymbol{#1}}}}

\newcommand{\tcm}[1]{{#1}}
\newcommand{\tcg}[1]{{#1}}

\def\apjl{The Astrophysical Journal Letter}

\def\aap{Astronomy \& Astrophysics}

\def\apjs{The Astrophysical Journal Supplement Series}

\shorttitle{\tcg{Local} Dark-Matter Map from Deep Learning}
\shortauthors{Hong et al.}
\begin{document}

\title{Revealing the Local Cosmic Web from Galaxies by Deep Learning}

\correspondingauthor{Donghui Jeong}
\email{djeong@psu.edu}

\author[0000-0003-4923-8485]{Sungwook E. Hong \begin{CJK*}{UTF8}{mj}(홍성욱)\end{CJK*}}
\affiliation{Natural Science Research Institute, University of Seoul, 163 Seoulsiripdaero, Dongdaemun-gu, Seoul, 02504, Republic of Korea}
\affiliation{Korea Astronomy and Space Science Institute, 776 Daedeokdae-ro, Yuseong-gu, Daejeon 34055, Republic of Korea}

\author[0000-0002-8434-979X]{Donghui Jeong}
\affiliation{Department of Astronomy and Astrophysics, and Institute for Gravitation and the Cosmos, The Pennsylvania State University, University Park, PA 16802, USA}

\author[0000-0003-3428-7612]{Ho Seong Hwang}
\affiliation{Astronomy Program, Department of Physics and Astronomy, Seoul National University, 1 Gwanak-ro, Gwanak-gu, Seoul 08826, Republic of Korea}
\affiliation{Korea Astronomy and Space Science Institute, 776 Daedeokdae-ro, Yuseong-gu, Daejeon 34055, Republic of Korea}

\author[0000-0002-4391-2275]{Juhan Kim}
\affiliation{Center for Advanced Computation, Korea Institute for Advanced Study, 85 Heogiro, Dongdaemun-gu, Seoul, 02455, Republic of Korea}

\begin{abstract}
The $80\%$ of the matter in the Universe is in the form of dark matter that comprises the skeleton of the large-scale structure called the Cosmic Web. As the Cosmic Web dictates the motion of all matter in galaxies and inter-galactic media through gravity, knowing the distribution of dark matter is essential for studying the large-scale structure. 
\tcm{However, the Cosmic Web's detailed structure is unknown because it is dominated by dark matter and warm-hot inter-galactic media, both of which are hard to trace.} Here we show that we can reconstruct the Cosmic Web from the galaxy distribution using the convolutional-neural-network\tcg{-}based deep-learning algorithm. We find the mapping between the position and velocity of galaxies and the Cosmic Web using the results of the state-of-the-art cosmological galaxy simulations, \emph{Illustris-TNG}. We confirm the mapping by applying it to the \emph{EAGLE} simulation. Finally, using the local galaxy sample from \emph{Cosmicflows-3}, we find the dark-matter map in the local Universe. We anticipate that the local dark-matter map will illuminate the studies of \tcg{the} nature of dark matter and the formation and evolution of the Local Group. High-resolution simulations and precise distance measurements to local galaxies will improve the accuracy of the dark-matter map.
\end{abstract}

\keywords{Local Group; dark matter; large-scale structure of universe}

\section{Introduction}
Since Fritz Zwicky inferred its existence from the large velocity dispersion of the Coma cluster \citep{zwicky1933} and Vera Rubin confirmed it with the flat rotation curve of galaxies \citep{rubin1970}, astronomers have been only strengthening the necessity of the non-baryonic matter providing excess gravity. We call that dark matter. The most substantial pieces of evidence include an excessive mass-to-light ratio in the dwarf galaxies \citep{Aaronson:1983}, the mismatch between the X-ray map (gas distribution) and the weak gravitational lensing map \citep[mass distribution;][]{Clowe/etal:2006}, as well as the disparity between the heights of even- and odd-acoustic peaks in the temperature power spectrum of the cosmic microwave background \citep{wmap7}. Dark matter is also an indispensable component of the concordance cosmological model. Accounting for the measured expansion rate of the Universe \citep{planck2018} requires the matter component whose energy density is over five times larger than that of atoms for which the robust upper limit comes from big-bang nucleosynthesis \citep{cooke/etal:2014}. The observed large-scale distribution of galaxies \citep{sloan} and the map of weak-gravitational lensing potential \citep{abbott2018} also require the dark matter providing the skeleton of the large-scale structure within which \tcg{clouds of} atoms collapse to form stars and galaxies \citep{davis/etal:1985}.

With the essential role that dark matter plays in modern astronomy and cosmology, in the past few decades, there have been continuous efforts \tcg{to search} %for searching 
for the nature of dark-matter {\it particles} in the particle accelerators \citep{Atlas:2019,CMS:2019}, cosmic rays \citep{2015JCAP...09..023G}, gamma-rays \citep{2015PhRvL.115w1301A}, and high-energy neutrinos \citep{Icecube:2018}. Beyond the Milky-way halo, there have also been recent studies focusing on the dark-matter signals from the extra-galactic sources by cross-correlating the high-energy cosmic rays with the distribution of galaxies \citep{2016PhRvD..94l3005F,2020arXiv200206234F} and dark matter traced by weak-gravitational lensing \citep{2017MNRAS.467.2706T,FermiXDES:2019}. All searches for the dark matter particles thus far, however, have not concluded with a firm detection. They have been only narrowing down the possible dark-matter masses and the interaction strengths among dark matter particles as well as between dark matter and atoms \citep{2017PhRvL.118b1303A,2018EPJC...78..203A}. For these efforts of searching for the nature of dark matter, the most basic information currently lacking is the distribution of the dark matter, or Cosmic Web, in the local large-scale structure beyond the Milky-way halo. Of course, we have a good reason to believe that dark-matter halos surround each galaxy in the Universe. It is, however, also well known that the galaxies are biased, rather than faithful, tracers of the large-scale structure \citep{desjacques/etal:2018}.

In this article, we shall present a novel method of unveiling the Cosmic Web in the local Universe. As dark matter is dark, of course, we cannot observe them directly from the telescope. The only guaranteed way of searching for the dark matter is the same method for their discovery, through their gravitational influence on visible objects. On the inter-galactic scales, 
\tcg{dark matter} dominates the gravitational interaction and determines the cosmic velocity flow. We can, therefore, infer the distribution of dark matter by carefully studying the distribution and motion of galaxies. Taking the observed distribution of galaxies and their peculiar velocity flow, in what follows, we shall decipher the dark matter distribution, or Cosmic Web, within local $\sim 20\Mpch$.

When reconstructing the local dark-matter distribution directly from observed galaxy distributions, we face the following challenges. First, the local galaxy distribution at the low Galactic latitudes is hidden behind the intense radiation from the Galactic disk and contaminated by the interstellar gas and dust, which makes it hard to obtain the complete map of the galaxy distribution. Second, even if we had the complete map of galaxies, they are biased tracers of the large-scale structure; that is, the distribution of galaxies does not necessarily reflect the distribution of dark matter. 

Previous attempts \citep{gottloeber2010,libeskind2010,carrick2015,lavaux2016,carlesi2016} of making the local dark-matter map, therefore, have relied on the cosmological simulations constrained by the smoothed density field at high-Galactic latitudes. 
\tcg{Typically, a smoothing scale of a few Mpc is employed when matching the simulation output to the observation.} 
However, this observational constraint for the fully evolved galaxy distribution is non-trivial to implement because the simulation needs the density distribution at the initial time.
\tcg{Alternatively, the \emph{Bayesian Origin Reconstruction from Galaxies} \citep[BORG; see, e.g.,][]{jasche2013,jasche2015} approach uses the multiple Gaussian processes to draw the probability distribution of the initial density perturbation from a given galaxy distribution. As based on the dark-matter density field evolution by second-order Lagrangian perturbation theory (2LPT) and linear galaxy bias model, the method is also limited to, again, the scale larger than a few Mpc where the 2LPT and linear bias models are accurate.}
 
Here we overcome the challenges by taking a novel approach based on deep learning (DL)\tcg{. 
DL, as well as a conventional machine learning technique, has been introduced to measure the dark matter distribution from weak gravitational lensing or spatial distribution of dark matter halos \citep[e.g.,][]{modi2018,shirasaki2019,jeffrey2020}. 
On the contrary, our DL approach aims to reconstruct the local dark-matter map down to an Mpc-scale by incorporating}
{\it all} information in the observed galaxy data: the spatial distribution and the radial peculiar velocity of galaxies. 
We use the convolutional neural network (CNN)-based DL algorithm to find the mapping between the local dark-matter distribution and the observed positions and the radial peculiar velocities of local galaxies.

The structure of this paper is as follows.
In Section~\ref{sec:data}, we describe the simulation and observational data used for DL training and prediction, respectively.
In Section~\ref{sec:method}, we will briefly describe our DL architecture and the evaluation of our DL model.
In Section~\ref{sec:result}, we will show the reconstructed local dark matter map and its statistical robustness.
We will summarize our result in Section~\ref{sec:discuss}.

Throughout the paper, we assume a standard $\Lambda$CDM cosmology in concordance with the \emph{Planck} 2018 analysis \citep{planck2018}: $(\Omega_{\rm m}^0, \Omega_\Lambda^0, h) = (0.31, 0.69, 0.6777)$. 
It is similar to the standard cosmologies adopted in \emph{Illustris-TNG} and \emph{EAGLE} simulations: $(\Omega_{\rm m}^0, \Omega_\Lambda^0, h) = (0.3089, 0.6911, 0.6774)$ and $(0.307, 0.693, 0.6777)$, respectively \citep{springel2018, schaye2015}.

\section{Data}\label{sec:data}
\subsection{Observational Data: Cosmicflows-3}
We use the \emph{Cosmicflows-3} galaxy catalog \citep[][\textsf{CF3} hereafter]{tully2016}, one of the most comprehensive galaxy catalog\tcg{s} that provide distance, radial peculiar velocity, and luminosity of 17,647 galaxies up to $200~ {\rm Mpc}$. To produce a fair galaxy sample over the given region, we make the volume-limited sub-sample of the \textsf{CF3} as follows. First, as the number density of the \textsf{CF3} galaxies close to the Galactic plane (Galactic latitude $|b| < 10^\circ$) is lower than average, we only use the galaxies at $|b| > 10^\circ$. Also, we use the $B$-band absolute magnitude ($M_B$) compiled from \emph{Lyon Extragalactic Database} \citep[\textsf{LEDA};][]{paturel2003} as a proxy of the stellar mass \citep[$M_\star$;][]{wilman2012}\tcg{.} We set the $B$-band magnitude $-15$ as the selection criterion, which is sufficient for covering the $20\Mpch$- and $40\Mpch$-cubic volume around the Milky-way galaxy. We have also tested the cases with $M_B < -16$ and $-17$ and \tcg{found} no noticeable difference \tcg{of} 
the predictions from the fiducial choice \tcg{(see Section~\ref{sec:result})}. Note that we have not used the $K_S$-band absolute magnitude, one of the best-known tracer\tcg{s} of the stellar mass \citep{bell2003}  
because that information is missing for about 30\% of the galaxies in our sample \citep{lavaux2011,huchra2012}.

We calculate the radial peculiar velocity by subtracting the Hubble flow from the velocity in the Galactic Standard of Rest \citep[$V_{\rm GSR}$;][]{kourkchi2020}. Note that we do not use the velocity in the \tcg{cosmic microwave background} (CMB) standard of rest ($V_{\rm CMB}$)%, 
to reduce any bias that might be introduced in the conversion. Instead, when generating training and test samples from simulation data, we include the peculiar motion of the Milky-way corresponding galaxy in each simulation. There exists a difference on the Hubble constant between recent CMB observations \citep[$H_0 = 67.77\,{\rm km/s/Mpc}$;][]{planck2018} and the best-fit from the \textsf{CF3} \citep[$H_0 = 75\,{\rm km/s/Mpc}$;][]{tully2016}. 
In this study, we have tested both values and find that the effect from the different Hubble constants stays within the uncertainty of the dark-matter map \tcg{(see Section~\ref{sec:result})}.

\subsection{Simulation Data: Illustris-TNG \& EAGLE}
We use the \textsf{TNG100-1}, a simulation with \tcg{a comoving volume} $V = (75\Mpch)^3$ and $1820^3$ dark-matter and gas particles from the \emph{Illustris-TNG} simulation suite \citep{springel2018,pillepich2018,marinacci2018,naiman2018,nelson2018,nelson2019}, as our high-resolution simulation data (\textsf{TNG100} hereafter). To mimic the observation from the Milky-way galaxy, we select 988 galaxies with stellar mass 
$4 \times 10^{10} \Msun < M_\star < 10^{11} \Msun$
(\emph{center galaxies} hereafter)%, 
by adopting that the Galactic stellar mass is about $5.2 \times 10^{10} \Msun$ \citep{licquia2015}. Around each center galaxy, we make a sub-cube with $20\Mpch$ box-size and calculate the dark-matter density field within the $64^3$ uniform grid. We also calculate the relative position of galaxies with $M_B < -15$ (\emph{target galaxies} hereafter) 
\tcg{and} the difference of peculiar velocity between the target galaxy and center galaxy.

For the low-resolution dark-matter map with $V = (40\Mpch)^3$, we use the \textsf{TNG300-1} from the \emph{Illustris-TNG} simulations, whose volume and number of particles are $V = (205\Mpch)^3$ and $2500^3$, respectively (\textsf{TNG300} hereafter). Note that the amplitude of the luminosity function of \textsf{TNG300} is lower than the observation and \textsf{TNG100}, mainly due to the lower spatial resolution of the simulation \citep{pillepich2018}. Therefore, we \tcg{also} apply the \emph{resolution correction} to find the center and target galaxies using the number density obtained from \textsf{TNG100} rather than directly using the face values of $M_\star$ or $M_B$. We also use the \textsf{TNG300-1-Dark}, a dark-matter-only counterpart of the \textsf{TNG300}, to test how baryonic physics affects our result. We select the center and target galaxies by finding the mass cut of dark matter halos with the same number density. The result from the \textsf{TNG300-1-Dark} is similar to or slightly worse than \textsf{TNG300} \tcg{(see Section~\ref{sec:result})}.

Also, we use the \textsf{RefL0100N1504}, a reference simulation with $V = (67.77\Mpch)^3$ and $1504^3$ dark-matter and gas particles from the \emph{EAGLE} simulation suite \citep[][\textsf{EAGLE} hereafter]{schaye2015,crain2015}, to check the \tcg{fidelity} of our result. For the center galaxies, we use the same selection criterion to \textsf{TNG100} and find 478 center galaxies. For the target galaxies, however, we do not directly use $M_B$. It is because the luminosity function of \textsf{EAGLE} is reliable only for bright galaxies ($M_B \lesssim -18$) since the \emph{EAGLE} simulations calculate the luminosity only to massive galaxies \citep[$M_\star \geq 10^{8.5} {\rm M_\odot}$;][]{camps2018}. Instead, similar to \textsf{TNG300}, we use the galaxy number density obtained from \textsf{TNG100} to find the stellar mass cut of target galaxies.

\section{Methods}\label{sec:method}

\subsection{Deep Learning Architecture}
\begin{figure*}[hbt]
\plotone{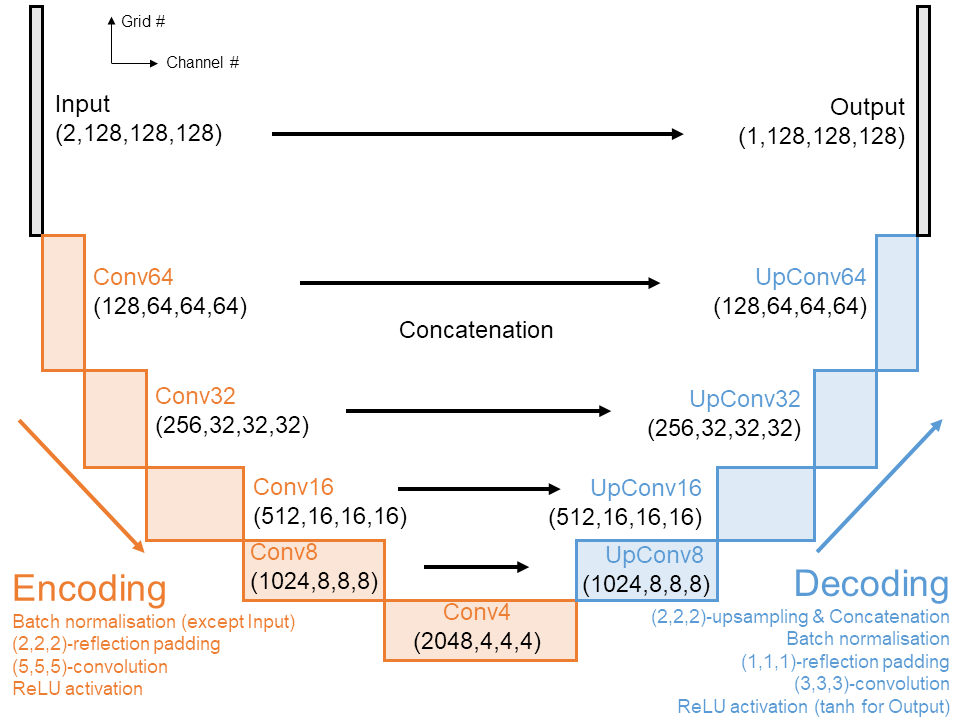}
\caption{The convolutional neural network (CNN) architecture used for \textsf{TNG300}. \tcg{We denote the layer size by the quadruple where the spatial dimension ($2^n, 2^n, 2^n$) follows the number of channels.} 
The size (except the number of filters) of each layer for \textsf{TNG100} is the half of \textsf{TNG300}.
\label{fig:architecture}}
\end{figure*}

\tcg{We construct the deep learning architecture using convolutional neural network (CNN) that highlights features in the data by a series of convolutions, resulting in so-called hidden layers. By varying the convolution filters, one can extract different physical features in the data. Specifically, we use a CNN architecture similar to the U-Net \citep{olaf2015} or V-Net \citep{milletari2016} to predict the dark-matter density field from the galaxy position and radial peculiar velocity (see Figure~\ref{fig:architecture}). Our CNN architecture consists of the following two stages: the encoding stage (\textsf{Input--Conv$N_{\rm s}$}) with increasing number of filters and decreasing the size of hidden layers, and the decoding stage (\textsf{UpConv$N_{\rm s}$--Output}) with decreasing number of filters and increasing the size of hidden layers. Here, $N_{\rm s}$ denotes the spatial size of hidden layers. To retain the small-scale spatial resolution, we also attach the hidden layers in the equivalent (with the same layer size) encoding stage as additional channels to the decoding layer, doubling the number of channels. We refer this process as concatenation.

The encoding stage consists of a series of 
\textsf{Conv}$N_{\rm s}$ layers. Let us define the input of a given \textsf{Conv}$N_{\rm s,0}$ as $\mathbb{I}_{\ell;i,j,k}$, where $i,j,k \in \left[1,N_{\rm s,0}\right]$ is the spatial coordinates, and $\ell \in \left[1, N_{\rm ch,0} \right]$ is the channel index with $N_{\rm ch,0}$ being the total number of channels. To accommodate the convolution at the edge, we have added the buffer around the input array (\emph{padding} process). As we use $5\times5\times5$ convolution filter, it suffices to add $N_{\rm p}=2$ padding pixels at both edges of each dimension. We fill the padding pixels by reflecting the inner two pixels next to the edge pixels.

After the padding, we apply a three-dimensional convolution with a multi-channel filter $w_{\ell,\ell';i',j',k'}$ and bias $b_\ell$, with 
indices 
$i',j',k' \in [1,N_{\rm k}=5]$, 
$\ell \in [1,N_{\rm ch,1}]$, 
$\ell' \in [1,N_{\rm ch,0}]$), to obtain the output ${\mathbb{C}}$ as
\begin{equation}
\mathbb{C}_{\ell;i,j,k} = b_\ell + \sum_{\ell',i',j',k'} \mathbb{P}_{\ell';s(i';i),s(j';j),s(k';k)} w_{\ell,\ell';i',j',k'} \, ,
\end{equation}
where ${\mathbb{P}}_{\ell';s(i';i),s(j';j),s(k';k)}$ is the input array after the padding.
We sample the convolution sparsely $s(i';i) = i\times N_{\rm st} + i'$, and reduce the spatial dimension by a factor of $2^3$ at each step by choosing the spatial interval $N_{\rm st}=2$ ({\it strides} hereafter). Accompanying the reduction of spatial dimension, we increase the number of channels $N_{\rm ch}$ by a factor of 2 at each step of the convolution, from 128 (\textsf{Conv64}) to 2048 (\textsf{Conv4}). Note that the convolution filter $w_{\ell,\ell';i',j',k'}$ and bias $b_\ell$ are \emph{trainable parameters} which we adjust for the training.

The padding and convolution processes are linear operations, so any combinations of these operations simplify to a single linear algebra operation. In order to fully utilize the multiple hidden layers of Deep Learning, we apply the rectified linear unit \citep[ReLU;][]{hahnloser2000, glorot2011},
\begin{equation}
\mathbb{A}_{\ell;i,j,k} = \max \left( \mathbb{C}_{\ell;i,j,k}, 0 \right) \,,
\end{equation}
as a non-linear \emph{activation function} for each hidden layer. 

Finally, we apply the batch normalization \citep{ioffe2015} 
\begin{equation}
\mathbb{O}_{\ell;i,j,k} = \gamma_{\ell;i,j,k} 
\frac{\mathbb{A}_{\ell;i,j,k} - \mu_{\ell;i,j,k}}{\sigma^2_{\ell;i,j,k} + \epsilon} + \beta_{\ell;i,j,k} \, ,
\end{equation}
to obtain an output \textsf{Conv}$N_{\rm s,1}$ layer, $\mathbb{O}_{\ell;i,j,k}$ ($i,j,k \in [1,N_{\rm s,1} = N_{\rm s,0}/2]$, $\ell \in [1,N_{\rm ch,1}]$).
Here, $\mu_{\ell;i,j,k}$ and $\sigma_{\ell;i,j,k}$ are the mean and standard deviation of $\mathbb{A}_{\ell;i,j,k}$ over samples in a same \emph{mini-batch}, and $\epsilon = 10^{-3}$ is a small value for the numerical stability. 
Note that the mini-batch refers to the bundle of input-output pairs that we have used for updating the trainable parameters.
The normalization factor $\gamma_{\ell;i,j,k}$ and bias factor $\beta_{\ell;i,j,k}$ are another trainable parameters. The batch normalization introduces extra level of non-linearity ensuring that the trainable parameters introduced at earlier hidden layers still affect the output.

The decoding stage consists of a series of \textsf{UpConv}$N_{\rm s}$ layers, which are constructed in a parallel manner.
In contrast to the \textsf{Conv}$N_{\rm s}$, 
where we decreases the spatial dimension by sparsely sampling the convolved array, 
we increase the spatial dimension of each \textsf{UpConv}$N_{\rm s}$ layer:
\begin{equation}
\mathbb{U}_{\ell;i,j,k} = \mathbb{I}_{\ell;u(i),u(j),u(k)},
\end{equation}
by duplicating the input array $\mathbb{I}_{\ell;i,j,k}$. 
Here, $u(x) = \lceil x / N_{\rm u} \rceil$, and we set the upsampling factor $N_{\rm u}=2$ in order to increase the spatial size of $\mathbb{U}_{\ell;i,j,k}$ by a factor of 8. 
After the upsampling, we concatenate the \textsf{Conv}$N_{\rm s}$ layer (the same size), and apply batch normalization.
We then apply a three-dimensional convolution with $(N_{\rm k},N_{\rm st}) = (3,1)$, 
after the reflective padding the edge arrays with $N_{\rm p} = 1$.
We decrease the number of output channels of each \textsf{UpConv$N_{\rm s}$} from 1024 to 128 by a factor of 2.
After the convolution, we apply the ReLU activation function.

In addition to the usual steps described above, the final {\textsf{Output}} layer requires following two special treatments so that the output layer represents the single dark-matter density proportional to ${\rm log}_{10}(\rho/\rho_0)$ which can be both positive and negative. First, instead of a gradual decrease of the number of output channels by a factor of 2, we set the number of output channels for \textsf{Output} as 1. Second, instead of the ReLU activation function, whose output range is $[0,+\inf)$, we use the hyperbolic tangent function ($\tanh$) so that its output range becomes finite ($[-1,+1]$ in this case).

We have adopted different spatial size of the hidden layer for \textsf{TNG100} and \textsf{TNG300} to accommodating the difference in their spatial resolution.
For \textsf{TNG100}, the encoding stage starts from $2$ channels of $64^3$-grid input layers, and ends with the 2,048 channels of the $2^3$-grid layer (\textsf{Conv2}), and, for \textsf{TNG300}, the encoding stage starts from $2$ channels of $128^3$-grid input layers, and ends with the 2,048 channels of the $4^3$-grid layer (\textsf{Conv2}).
The final output layers are $64^3$ and $128^3$ for, respectively, {\sf TNG100} and {\sf TNG300}. We have also tested other CNN architectures with various channel sizes, and confirmed that the CNN architecture that we use here (shown in Figure~\ref{fig:architecture}) performs the best among the tested cases.
}

\subsection{Training}

We divide the training and validation samples from \textsf{TNG100} so that all sub-cubes from the validation sample do not overlap with those from the training sample. As a result, we only use 525 sub-cubes --- 432 for training and 93 for validation. For each sub-cube, we make two $64^3$ uniform grids as a two-channel input layer; each channel stores the number of target galaxies \tcg{($N_{\rm gal}$)} and the averaged radial peculiar velocity \tcg{($V_{\rm pec}$)} in units of ${\rm km/s}$. For the input layer, we apply the same Galactic-latitude mask as the CF3 data (masking out $|b|<10^\circ$). For the output layer, we normalize the logarithm of dark-matter density to be 
\begin{equation}
y = \frac{1}{4.5} \log_{10} (\rho/\rho_0) \, \tcg{,}
\label{eq:def_y}
\end{equation}
\tcg{where $\rho_0$ is the mean dark matter density of the Universe} so that all values in the output layer would be between $-1$ and $+1$.

For data augmentation, we allow swapping the $(x,y,z)$-axes of each sub-cube, which increases the number of samples by a factor of three. We further increase the sample size by flipping the axis direction, with which the number of samples increases eight times. Note that, unlike U-Net or V-Net, we do not split a single cube into multiple smaller cubes for data augmentation %, 
because that would change of \tcg{the} Galactic-latitude mask and the radial peculiar velocity. In the end, we obtain 10,368 and 2,232 samples, respectively, in training and validation sets.

We \tcg{implement} our CNN architecture \tcg{ in } the \emph{Keras} \citep{chollet2015} with the \emph{Tensorflow} backend \citep{abadi2015} \tcg{and perform the training with the \emph{NVIDIA Tesla V100} graphic processing unit (GPU) with 16GB memory}. 
\tcg{We choose the mean squared error (MSE) as
the loss function that the DL minimizes during the training: 
\begin{align}
\mathcal{L}_{\mathsf{TNG100}} &= \frac{1}{n} \sum_{i=1}^{n} (y_{i,{\rm pred}} - y_{i,{\rm truth}})^2 \\
&= \frac{1}{n} \sum_{i=1}^{n} \left[ \frac{1}{4.5} \log_{10} (\rho_{i,{\rm pred}} / \rho_{i,{\rm truth}} ) \right]^2 \, ,
\end{align}
where the subscripts $(i,{\rm pred})$ and $(i,{\rm truth})$ are, respectively, the prediction and truth values of the 
$y$ (defined in Eq.~\ref{eq:def_y}) at $i$-th grid.}

\tcg{Initially, we set the trainable parameters in the convolution filters ($\vec{\theta}$; \emph{parameter vector} hereafter) randomly. The training process for minimizing the loss function is done with 200 \emph{epochs}, a unit process that updates the parameter vector from a subset of the train set and applies the updated parameter vector to a subset of the validation set. The parameter vector update process at each epoch consists of 1728 mini-batches. We set the mini-batch size as 6, mainly due to the GPU memory limit. 
For each mini-batch we numerically calculate  the gradient of the loss function ($\nabla_{\vec{\theta}} \mathcal{L}$) and update the parameter vector by the Adam optimizer \citep{kingma2014},
\begin{align}
\vec{\theta}_t &= \vec{\theta}_{t-1} - \alpha \frac{\vec{m}_t / (1-\beta_1^t)}{\sqrt{\vec{v}_t / (1-\beta_2^t)} + \epsilon} \\
\vec{m}_t &= \beta_1 \vec{m}_{t-1} + (1-\beta_1) \nabla_{\vec{\theta}} \mathcal{L}_t(\vec{\theta}_{t-1}) \\
\vec{v}_t & = \beta_2 \vec{v}_{t-1} + (1-\beta_2) \left[ \nabla_{\vec{\theta}} \mathcal{L}_t(\vec{\theta}_{t-1}) \right]^2 \, .
\end{align}
Here $t$ is a mini-batch step number starting from zero, $\vec{m}_t$ and $\vec{v}_t$ are the first and second-moment vectors with initial values $\vec{m}_0 = \vec{v}_0 = \vec{0}$, $\beta_1 = 0.9$ and $\beta_2 = 0.999$ are exponential decay rates for moment estimates, and $\epsilon = 10^{-7}$ is a small value for the numerical stability. 
$\alpha$ is the \emph{learning rate} that determines how fast one updates the parameter vector, and we set it as $10^{-3}$. 
As a result, the training process for \textsf{TNG100}} takes about 73 hours for a single run.

\begin{figure}[hbt]
\plotone{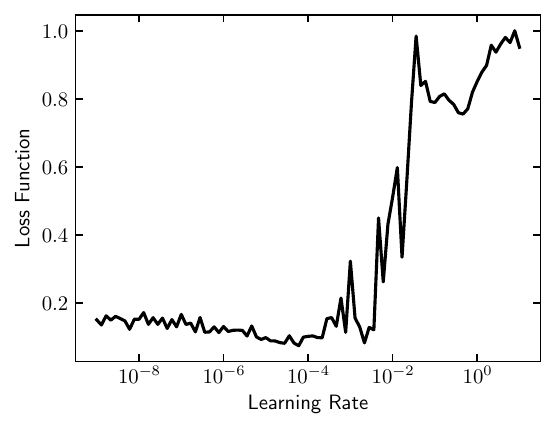}
\caption{\tcg{Evolution of loss function ($\mathcal{L}$) as a function of learning rate of Adam optimizer ($\alpha$) from an additional test training for \textsf{TNG300}.
Too low learning rate ($\alpha \lesssim 10^{-8}$) gives a too slow update of the parameter vector, which is presented as a flat slope of $\mathcal{L}(\alpha)$.
On the other hand, too high learning rate ($\alpha \gtrsim 10^{-5}$) prevents finding a solution, which is presented as a noisy increment of $\mathcal{L}(\alpha)$.}}
\label{fig:learning_rate_finding}
\end{figure}

We perform a similar training for the \textsf{TNG300} outcome, except \tcg{for} the following differences. First, we have 10629 training sub-cubes and 1256 validation sub-cubes, with each sub-cube having $128^3$-grids. Unlike \textsf{TNG100}, we do not apply further data augmentation, mainly due to the expensive computational cost from large CNN architecture size. Second, since the dynamic range of dark-matter density of \textsf{TNG300} is wider than \textsf{TNG100}, we use 
\begin{equation}
y = \frac{1}{5} \log_{10} (\rho/\rho_0)
\end{equation}
for the output layer instead. \tcg{As a result, the MSE loss function becomes
\begin{equation}
\mathcal{L}_{\mathsf{TNG300}} = \frac{1}{n} \sum_{i=1}^{n} \left[ \frac{1}{5} \log_{10} (\rho_{i,{\rm pred}} / \rho_{i,{\rm truth}} ) \right]^2 \, .
\end{equation}} Third, 
\tcg{instead of using a fixed learning rate, we apply a triangular cyclic learning rate \citep{smith2015},
\begin{align}
\alpha_t  = \alpha_{\rm L} + \frac{\alpha_{\rm U}-\alpha_{\rm L}}{T/2}
\times \min\left\{(t\,{\rm mod}\,T), T-(t\,{\rm mod}\,T) \right\}  \, ,
\end{align}
to avoid the training to be stuck in local minima.
Here $T$ is the number of mini-batches that consists a single learning rate cycle, and we set it as 8. 
$\alpha_{\rm L}$ and $\alpha_{\rm U}$ are the minimum and maximum values of the learning rates, respectively. 
To find a suitable range of learning rates, we have performed an additional test training with a few epochs by varying learning rates (see Figure~\ref{fig:learning_rate_finding}). 
If the learning rate is too low, i.e., if the parameter vector update is too slow, the loss function as a function of learning rate ($\mathcal{L}(\alpha)$) has a flat slope. 
On the other hand, if the learning rate is too high, i.e., if an interval of parameter vector update is too large to find a solution, $\mathcal{L}(\alpha)$ presents a noisy increment. 
We found that $3 \times 10^{-8} < \alpha < 4 \times 10^{-5}$ is a suitable range of the learning rate and set $(\alpha_{\rm L}, \alpha_{\rm U}) = (3 \times 10^{-8}, 4 \times 10^{-5})$ for the triangular cyclic learning rate accordingly. }
Finally, due to the large CNN architecture size, we use four \emph{NVIDIA Tesla V100} \tcg{ GPUs with 32GB memory per each, } with \tcg{a} mini-batch size 
\tcg{of} 8. For each training, we run 400 epochs by using only 157 mini-batches per epoch, and it takes about 90 hours for a single run.

\begin{figure}[hbt]
\plotone{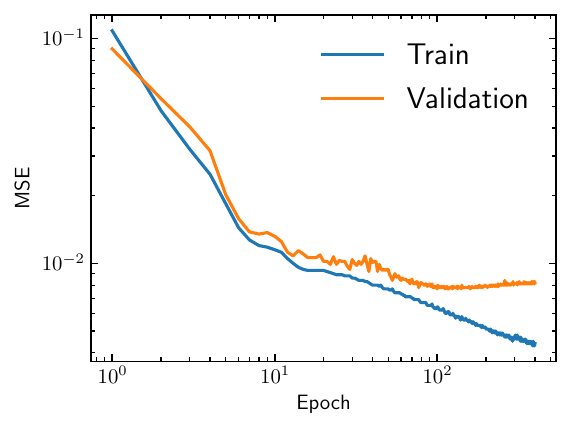}
\caption{\tcg{Evolution of loss functions from train (blue) and validation (orange) sets as a function of epoch for \textsf{TNG300}.}}
\label{fig:learning_curve}
\end{figure}

\tcg{Figure~\ref{fig:learning_curve} shows the evolution of the MSE loss functions from both train and validation sets as a function of epoch in \textsf{TNG300}.
Both train and validation losses similarly decrease over epoch until the validation loss reaches its minimum around $8 \times 10^{-3}$ at $\sim 140$ epochs, while the train loss continues decreasing at all epochs.
Similar minimum values of validation losses have been found during our test training, and we expect that the above value is close to the global minimum of the validation loss function in our current CNN setup.
If the validation loss greatly increases over epoch after reaching its minimum while the train loss keeps decreasing, it may infer that the learning process starts overfitting the data---the learning process tries to memorize the data without finding any global feature.
In our runs, however, the validation loss does not increase more than 1.1 times its minimum until the last epoch, which suggests that our result would not suffer from overfitting problem significantly.}
From each run, we select three models from three different epochs for the following performance test: at the minimum validation loss, at the minimum training loss, and the last epoch.

\begin{deluxetable*}{ll}[hbt]
\tablecaption{\tcg{Summary of \textsf{TNG300} and its comparison models used in this paper.
Each comparison model is the same as \textsf{TNG300} except those mentioned in its ``Description.''}}\label{tab:models}
\tablehead {
\colhead{Model} &
\colhead{Description}
}
\startdata
\textsf{TNG300} & Simulation: \textsf{TNG300-1} hydrodynamic simulation. \\
& Center galaxies: $4 \times 10^{10} {\rm M}_\odot < M_\star < 10^{11} {\rm M}_\odot$ after resolution correction.\\
& Target galaxies: $M_B < -15$ after resolution correction. \\
& Input layer: 2-channel ($N_{\rm gal}$ and $V_{\rm pec}$). \\
& Hubble parameter: $67.77 {\rm km/s/Mpc}$ \\
\hline
\textsf{16mag} & Target galaxies: $M_B < -16$ after resolution correction. \\
\textsf{17mag} & Target galaxies: $M_B < -17$ after resolution correction. \\
\textsf{noVpec} & Input layer: 1-channel ($N_{\rm gal}$). \\
\textsf{stellarMass} & Input layer: 2-channel ($\log_{10} (M_\star/{\rm M}_\odot)$ (logarithm of the total stellar mass) and $V_{\rm pec}$). \\
\textsf{DMhalo} &  Simulation: \textsf{TNG300-1-Dark} dark-matter-only simulation.\\
& Center \& target galaxies: applying halo mass cut that matches the same galaxy number density to \textsf{TNG300}.\\
\textsf{diffH0} & Hubble parameter: $75 {\rm km/s/Mpc}$ \\
\enddata
\end{deluxetable*}

For \textsf{TNG300}, we perform \tcg{six additional} alternative training \tcg{with different configurations of the input layer (\emph{comparison models} hereafter) to understand how such difference affects our prediction (see Table~\ref{tab:models})}. 
\tcg{\textsf{16mag} and \textsf{17mag} use the alternative absolute $B$-band magnitude cutoffs $M_B < -16$ and $-17$, respectively.
\textsf{stellarMass}} uses the logarithm of the total stellar mass rather than the simple galaxy number as an input layer,
\tcg{while \textsf{noVpec}} does not use the radial peculiar velocity 
\tcg{.
Finally, \textsf{DMhalo} uses the dark matter halos in the dark-matter-only simulation \textsf{TNG300-1-Dark} instead of galaxies in the \textsf{TNG300-1}.}\footnote{\tcg{Note that \citet{modi2018} performed a similar study to reconstruct the (initial) density perturbation from the dark matter halo distributions by DL, while they focused more on large scales such as baryon acoustic oscillation (BAO) rather than relatively small scales such as ours.}}

%%%%%%%%%%%%%%%%%%%%%%%%%%%%%%%%%%%%%%%%%%%%%%%%%%%%%%%%%%%
\section{Results}\label{sec:result}
\subsection{Performance Test}\label{sec:result_test}

\begin{deluxetable*}{lcccc}[hbt]
\tablecaption{\tcg{Summary of the performance test done by  validation samples of \textsf{TNG100}, \textsf{TNG300}, and their comparison models.
${\rm KS}(\xi_{\rm pred},\xi_{\rm truth})$ is the Kolmogorov-Smirnov statistics of the two-point correlation functions of dark-matter distribution between truth and prediction.
\textsf{EAGLE-TNG100} is the application of the \textsf{TNG100} model to the \textsf{EAGLE} samples.
\textsf{diffH0} is identical to \textsf{TNG300} since Hubble flow estimation is not considered in this test.}}\label{tab:perform}
\tablecolumns{5}
\tablehead {
\multirow{2}{*}{Model} &
\multirow{2}{*}{$\log_{10} (\rho_{\rm pred}/\rho_{\rm truth})$} &
\multicolumn{3}{c}{${\rm KS}(\xi_{\rm pred},\xi_{\rm truth})$} \\
& &
\colhead{$0-1\Mpch$} &
\colhead{$1-3\Mpch$} &
\colhead{$3-10\Mpch$}
}
\startdata
\textsf{TNG100}
& $-0.014 \pm 0.543$
& $0.263 \pm 0.035$ & $0.175 \pm 0.087$ & $0.130 \pm 0.042$\\
\textsf{EAGLE-TNG100}
& $+0.129 \pm 0.491$ & $0.171 \pm 0.055$ & $0.152 \pm 0.047$ & $0.149 \pm 0.040$ \\
\hline
\textsf{TNG300}
& $-0.020 \pm 0.451$ & $0.153 \pm 0.035$ & $0.134 \pm 0.040$ & $0.163 \pm 0.017$ \\
\textsf{16mag}
& $-0.008 \pm 0.468$ & $0.109 \pm 0.010$ & $0.161 \pm 0.033$ & $0.254 \pm 0.016$ \\
\textsf{17mag}
& $+0.017 \pm 0.481$ & $0.143 \pm 0.037$ & $0.168 \pm 0.018$ & $0.251 \pm 0.019$ \\
\textsf{noVpec}
& $+0.016 \pm 0.481$ & $0.367 \pm 0.115$ & $0.407 \pm 0.061$ & $0.170 \pm 0.036$ \\
\textsf{stellarMass}
& $-0.050 \pm 0.471$ & $0.186 \pm 0.056$ & $0.218 \pm 0.016$ & $0.269 \pm 0.021$ \\
\textsf{DMhalo}
& $+0.002 \pm 0.481$ & $0.264 \pm 0.029$ & $0.243 \pm 0.030$ & $0.263 \pm 0.034$ \\
\enddata
\end{deluxetable*}

\begin{figure*}[hbt]
\includegraphics[width=\textwidth]{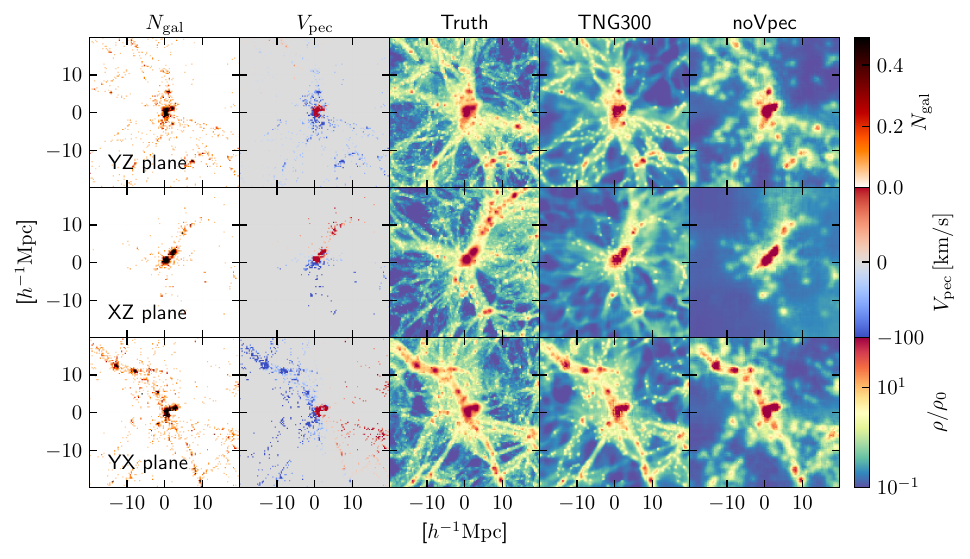}
\caption{3-way projections of a single \textsf{TNG300} validation sample with $5\Mpch$-thickness.
From left to right: galaxy number \tcg{($N_{\rm gal}$)}, radial peculiar velocity \tcg{($V_{\rm pec}$)}, \tcg{truth} dark-matter density \tcg{($\rho_{\rm truth}$)}, reconstructed dark-matter density \tcg{($\rho_{\mathsf{TNG300}}$), and another reconstruction from the CNN architecture without using the radial peculiar velocity (\textsf{noVpec}; $\rho_{\mathsf{noVpec}}$).
\textsf{TNG300} can well reconstruct the filamentary structure of a few-Mpc scales in the true dark-matter distribution, while \textsf{noVpec} does not show such structure}.
\label{fig:DLvisual}}
\end{figure*}

\begin{figure*}[hbt]
\centering
\includegraphics[width=0.329\textwidth]{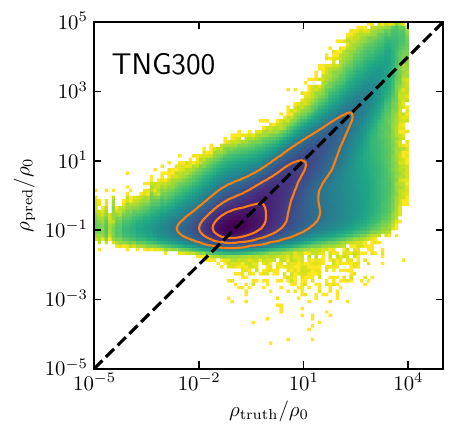}
\includegraphics[width=0.329\textwidth]{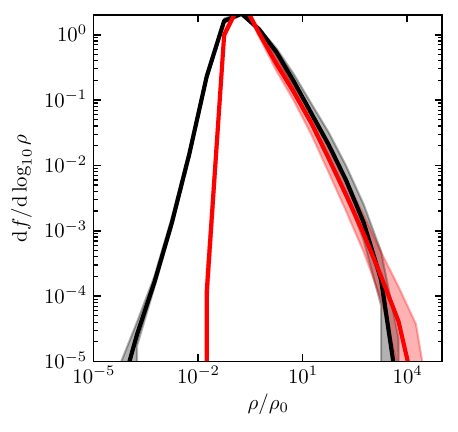}
\includegraphics[width=0.329\textwidth]{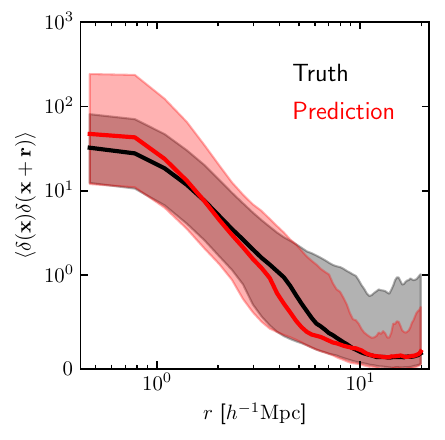} \\
\includegraphics[width=0.329\textwidth]{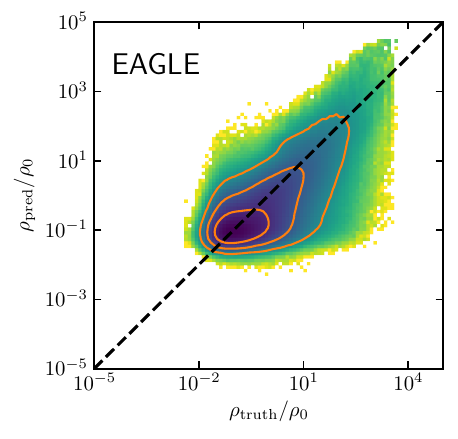}
\includegraphics[width=0.329\textwidth]{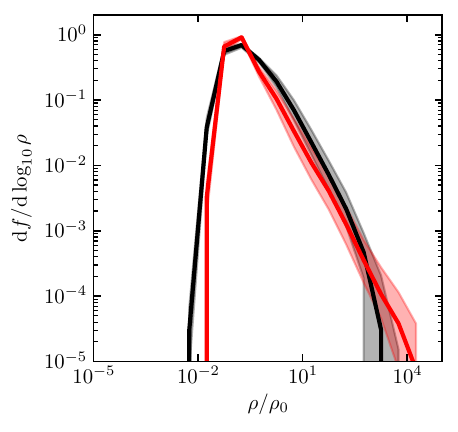}
\includegraphics[width=0.329\textwidth]{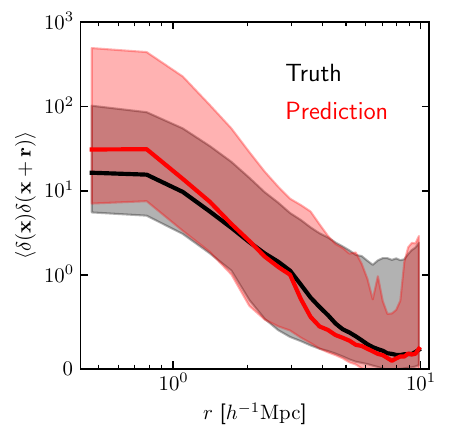}
\caption{Result of the performance tests for the deep learning result using the three-dimensional dark-matter density field of simulations.
Top panel: statistical comparison between the ground truth and the predicted dark-matter density from the entire \textsf{TNG300} validation sample.
From left to right: joint probability distribution (colors) with $1,2,3$-$\sigma$ certainty level contours (lines), median (lines) and 1-$\sigma$ deviation (shades) of histograms, and median (lines) and 1-$\sigma$ deviation (shades) of the two-point correlation functions.
Bottom panel: similar to the top panel, but by applying the \textsf{TNG100} training to the entire \textsf{EAGLE} test sample.
\label{fig:DLresult}}
\end{figure*}

\tcg{To test the model parameters tuned with \textsf{TNG100} and \textsf{TNG300} training sets, we apply the model to the validation samples to} 
compare the resulting dark-matter density cube with the ground truth. Specifically, we use the following four methods for the performance test---visual comparison, joint probability distribution, histogram, and two-point correlation function (2pCF) \tcg{$\xi(r) = \langle \delta(\mathbf{x}) \delta(\mathbf{x}+\mathbf{r}) \rangle_{\mathbf{x}}$} 
. To examine the performance of the each model, we use \tcg{the Kolmogorov-Smirnov statistics of the} 2pCF\tcg{s between truth and prediction at a given scale,
\begin{equation}
{\rm KS}(\xi_{\rm pred},\xi_{\rm truth}) = \max_{\xi'} |\tilde{P}_{\rm pred}(\xi') - \tilde{P}_{\rm truth}(\xi')| \, .
\end{equation}
Here, $\tilde{P}(\xi') = N(\xi < \xi')/N_{\rm sample}$ is the empirical distribution function}%.}
\tcm{, where $N_{\rm sample}$ and $N(\xi < \xi')$ are, respectively, the number of whole samples and the number of those satisfying $\xi < \xi'$.
The smaller ${\rm KS}(\xi_{\rm pred}, \xi_{\rm truth})$ indicates that the predicted probability distribution of the 2pCF is closer to the true distribution, so we use that as a metric to compare} the performance of models. For both \textsf{TNG100} and \textsf{TNG300}, the models at the minimum training loss provide the closest distribution of the 2pCF predictions to their truth, and we adopt them as our optimal models.

\tcg{Table~\ref{tab:perform}, Figures~\ref{fig:DLvisual} and \ref{fig:DLresult} show a visual inspection and the statistics of the \textsf{TNG300} validation samples, which show a good agreement with their true values.
Interestingly, the predicted dark matter distribution shows small-scale filamentary structures, which are not apparently shown in $N_{\rm gal}$ alone.
This is the first indication of the importance of the (radial) peculiar velocity field for reconstructing the small-scale filamentary structures; that is, the recovered dark-matter map shows much more detailed structure than simply connecting the galaxy positions, since the peculiar velocity could provide information about the underlying gravitational potential. Simply put, we use the galaxies as test particles for recovering the local gravitational field.
Note that, however, there is a slight difference in the detailed distribution of filamentary structures between truth and prediction.
Also, note that there exists a sharp lower cut in the predicted density $\min \rho_{\rm pred} \sim 10^{-2} \rho_0$.
The above two issues could be overcome by using higher-resolution hydrodynamic simulations and observational data with more low-brightness galaxies.
Also, fine-tuned choices of loss function might help manage an issue about a slight difference of filamentary structures.
}

After choosing the optimal models, we perform the convergence test between models with different simulation resolutions and setups. First, we compare the local dark-matter density field predictions from \textsf{TNG100} and \textsf{TNG300} within the radius $r = 10\Mpch$. We find that they show similar distribution up to $r \sim 4\Mpch$, while the dark-matter map from \textsf{TNG100} shows finer small-scale structures than \textsf{TNG300} (see \tcg{Section~\ref{sec:result_2d}}). Also, we apply the CNN model from \textsf{TNG100} to the test sample of \textsf{EAGLE} \tcg{(\textsf{EAGLE-TNG100} in Table~\ref{tab:perform})}\tcg{.} 

Note that we do not apply the CNN model from \textsf{TNG300} to \textsf{EAGLE} because the volume of \textsf{EAGLE} is not sufficiently larger than the volume of \textsf{TNG300} sub-cubes.
\tcg{We} find that its performance test result is similar to the \textsf{TNG100} validation sample\tcg{, except that \textsf{EAGLE-TNG100} tends to slightly overestimate the dark-matter density} (see Figure~\ref{fig:DLresult} \tcg{and Table~\ref{tab:perform}}).

\tcg{We also test the performance of various comparison models of \textsf{TNG300} (see Table~\ref{tab:models} for definitions).
Most comparison models show similar overall performance to \textsf{TNG300}, while those from the dark-matter-only simulation (\textsf{DMhalo}) has slightly more offset in the distribution of 2pCFs.
Those without using the radial peculiar velocity as inputs (\textsf{noVpec}), however, do not reproduce any small-scale filamentary structure shown in the true dark matter distribution (see the right panel of Figure~\ref{fig:DLvisual}).
From its visual inspection, one could interpret the output of \textsf{noVpec} as a smoothing of the galaxy number distribution---the only available input of the given DL model--- with a few Mpc-scale.
As a result, the 2pCFs of \textsf{noVpec} show a significant deviation from their truth in small scales with $r \lesssim 3 \Mpch$ (see Table~\ref{tab:perform}).
From the comparison to \textsf{TNG300} and its other comparison models, it is apparent that the (radial) peculiar velocity plays a significant role in reconstructing the small-scale filamentary structure.
}

\subsection{Three-dimensional view of the Local Cosmic Web}\label{sec:result_3d}

\begin{figure*}[hbt]
\centering
\begin{tabular}{cc}
\includegraphics[width=0.49\textwidth]{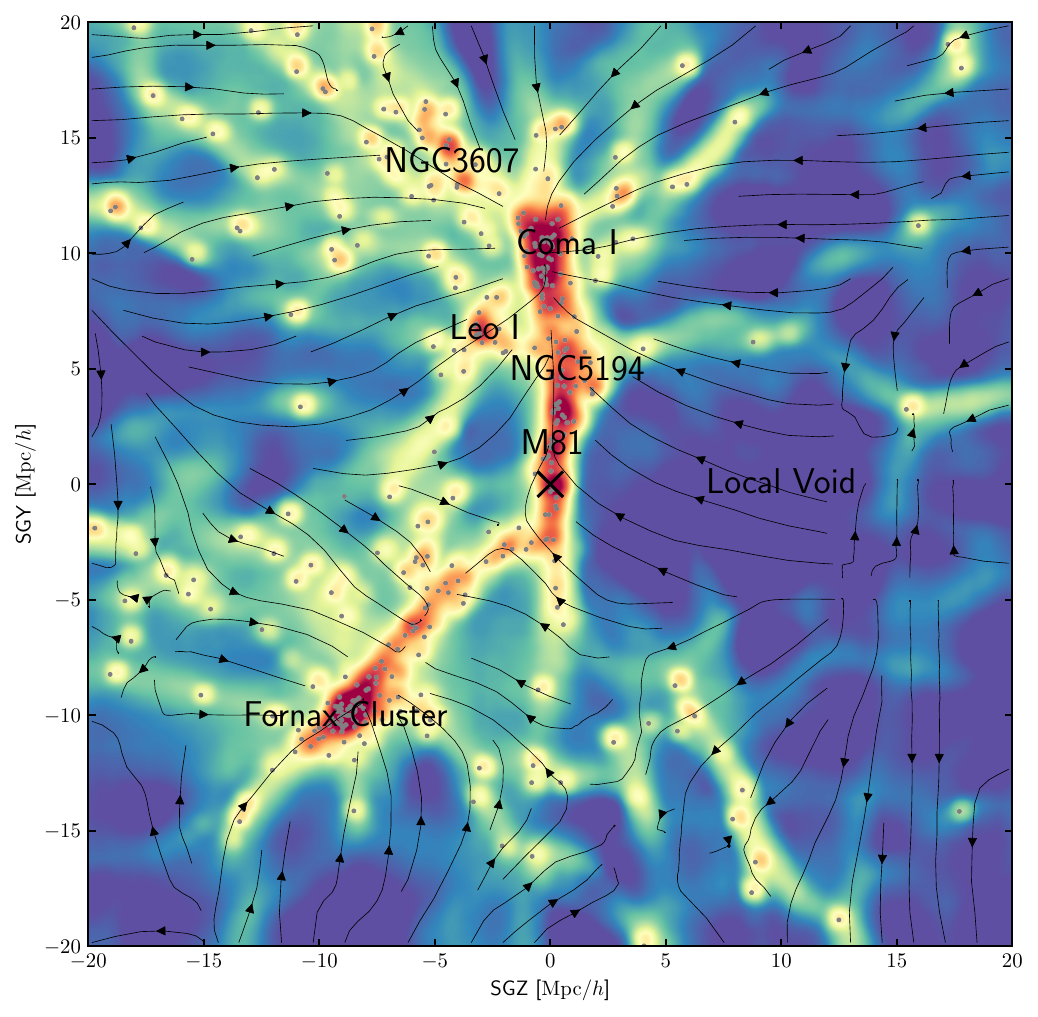} &
\\
\includegraphics[width=0.49\textwidth]{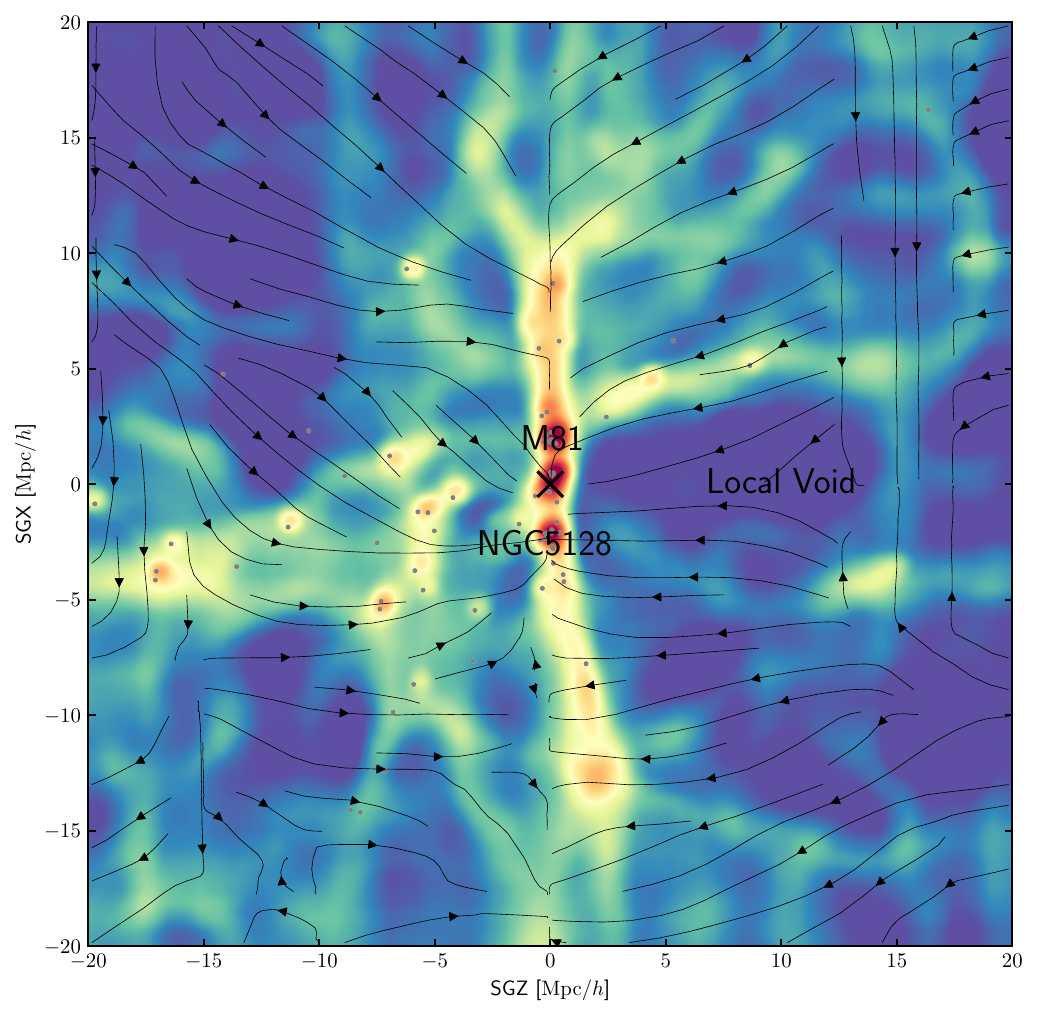} &
\includegraphics[width=0.49\textwidth]{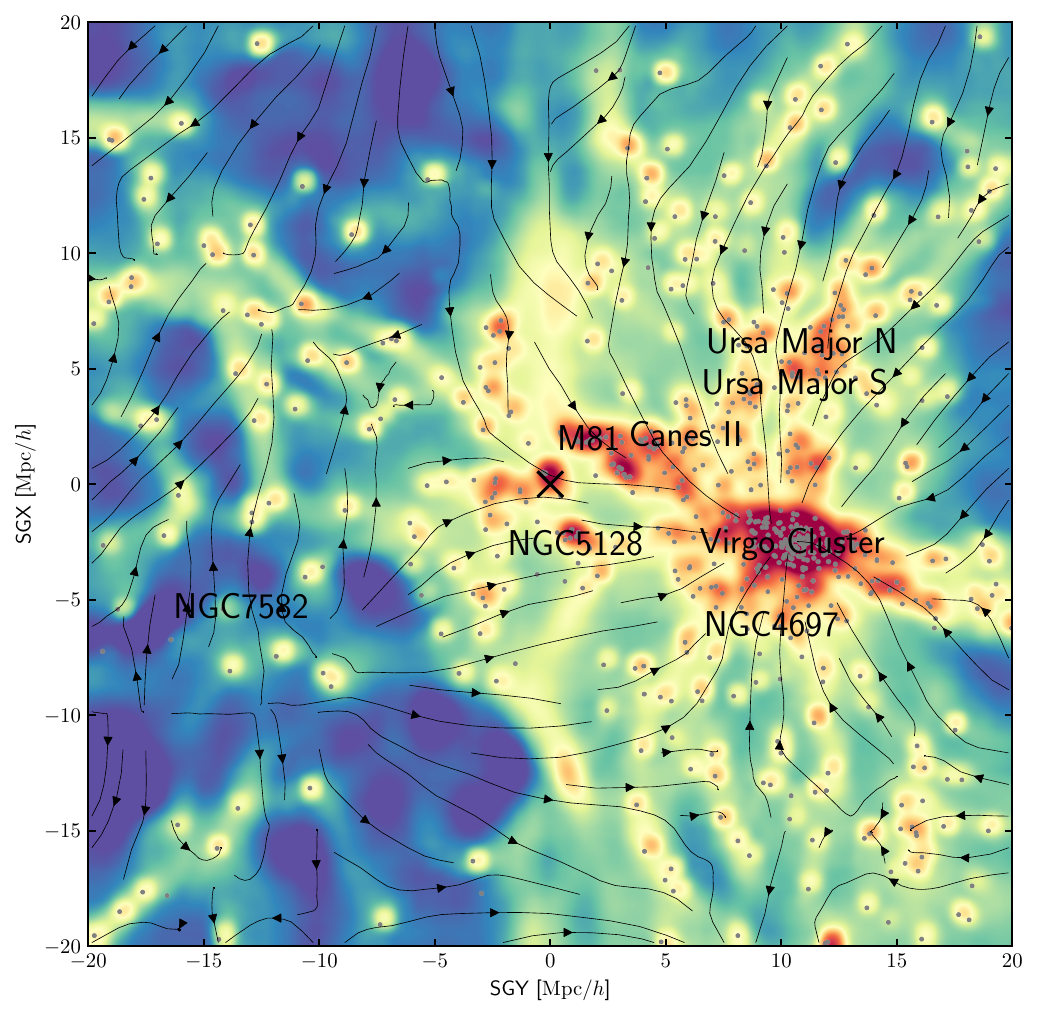} \\
\end{tabular}
\caption{Three-dimensional density maps of the local dark matter with $40\,{\rm Mpc}/h$-boxsize and $4\,{\rm Mpc}/h$-thickness.
`X'-mark at the center: Milky-way galaxy. 
Dots: galaxies with $M_B < -15$.
Texts: galaxy groups, clusters, and local structures.
Arrows: estimated directions of motion derived from the gradient of the reconstructed gravitational potential.
\label{fig:3DMap}}
\end{figure*}

Figure~\ref{fig:3DMap} shows a sliced view of the reconstructed Cosmic Web integrated over $4\Mpch$ thickness. Each panel shows the Cosmic Web on the plane of the Supergalactic Cartesian coordinates (SGX, SGY, and SGZ), extended to the full cube with the side length \tcg{of} $40\Mpch$. Figure~\ref{fig:3DMap} clearly shows known local objects that we designated by their common name. The figure also recovers known local large-scale structures. For example, we find a $10\Mpch$-spread along $+{\rm SGY}$-direction in the SGZ-SGY (upper left panel) and SGY-SGX (lower-right panel) planes. This structure is known as Local Sheet\tcg{, which } connects the Local Group and Virgo cluster and \tcg{contains} M81, NGC5194, Canes II, and Coma I groups \citep{localsheet08,courtois2013}. We also find that, around the Local Group, the Local Sheet is connected to the Fornax Wall \citep{fairall1994}, which is a $20\Mpch$-sized spread along ($-$SGY,$-$SGZ)-direction, containing Fornax cluster, Eridanus cluster, and Dorado group as members (upper-left panel). At the opposite direction to the Fornax Wall on the SGZ-SGY plane, the Local Void \citep{localvoid87} is also apparent (also shown on the SGZ-SGX plane), which might extend beyond the boundary of our local universe sample. In Figure~\ref{fig:3DMap}, we also present the velocity flow lines derived from the reconstructed gravitational potential \tcg{gradient} with arrows and black lines. The velocity flow shows the motion of material from the Local Void to nearby filamentary structures and clusters such as the Local Sheet, Fornax Wall, and Virgo cluster. 
\tcg{Note} that we cannot reproduce the velocity flow from the Virgo cluster to the Great Attractor ($+{\rm SGX}$-direction), because of the limited extension of the volume that we analyze here. However, we would like to emphasize that the recovered dark-matter map provides us detailed density and velocity fields around these known local large-scale structures. 

The recovered Cosmic Web also shows a hint of new structures that require further investigation. For example, the direction of the Local Sheet is similar to the direction of the so-called vast polar structure (VPOS), which consists of satellite galaxies, globular clusters and stellar streams around the Milky-way galaxy \citep{pawlowski2012}. As shown in Figure~\ref{fig:3DMap}, the Local Sheet, being the strongest filamentary structure around the Local Group, is a source of velocity flow; that might cause a connection between the two. Also, a couple of small filaments \tcg{are} visible in our maps, which could be good targets for systematic examination with deep imaging surveys.

Furthermore, to estimate the uncertainties of the dark-matter map, we perform a stress test \tcg{on} our CNN models by incorporating distance measurement \tcg{uncertainties} in the \textsf{CF3}. 
We use the \tcg{one standard deviation} uncertainty \tcg{in} distance modulus \tcg{($\epsilon_\mu$) in the \textsf{CF3},
\begin{equation}
\epsilon_\mu \equiv \sqrt{ \frac{1}{\sum_i 1/\epsilon_i^2} } \, .
\end{equation}
Here $\epsilon_i$ includes the one standard deviation uncertainty determined from a recalibration of galaxy magnitude with \ion{H}{1} linewidth \citep{tully2012}, distance measurement of the Tip of the Red Giant Branch from the \emph{Hubble Space Telescope}, Type Ia supernovae from various samples \citep{tully2013}, Tully-Fisher relation using \emph{Spitzer} $[3.6]$ photometry, and the Fundamental Plane relation from the \emph{Six Degree Field Galaxy Survey} (\emph{6dFGS}) \citep{tully2016}.
We then} generate 1,000 sets of random distance moduli that follow the normal distribution\tcg{,
\begin{equation}
P(\Delta \mu) = \frac{1}{\epsilon_{\mu} \sqrt{2 \pi}} \exp \left[ -\frac{\Delta \mu^2}{2 \epsilon_{\mu}^2} \right] \, .
\end{equation}}
Then, we re-calculate the radial peculiar velocity by subtracting the Hubble flow corresponding to the random distances from the $V_{\rm GSR}$. Since the distance measurement error exists only along the radial direction, we have generated the two-dimensional column density map of the dark matter that is less affected by the error than the three-dimensional dark-matter density field (see Figure~\ref{fig:StatMap}). Also, we find that the dark-matter column density map driven from \textsf{TNG300} shows significantly less deviation than that of \textsf{TNG100}, which suffers from some spurious structure consistently appearing near the Galactic plane.

\subsection{Sky map of the Local Cosmic Web}\label{sec:result_2d}

\begin{figure*}[hbt]
\centering
\begin{tabular}{c|c}
\includegraphics[width=0.35\textwidth]{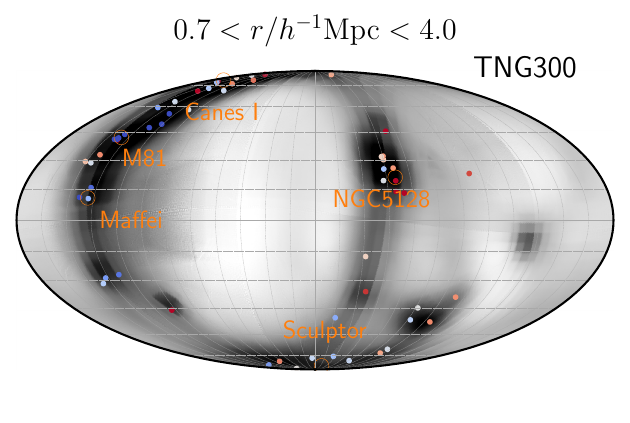}
\hspace{0.025\textwidth} &
\hspace{0.025\textwidth}
\includegraphics[width=0.35\textwidth]{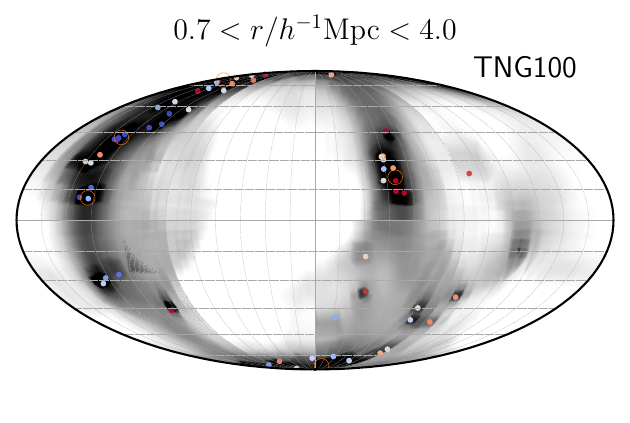}
\vspace{-15pt}\\ \cline{2-2}
\includegraphics[width=0.35\textwidth]{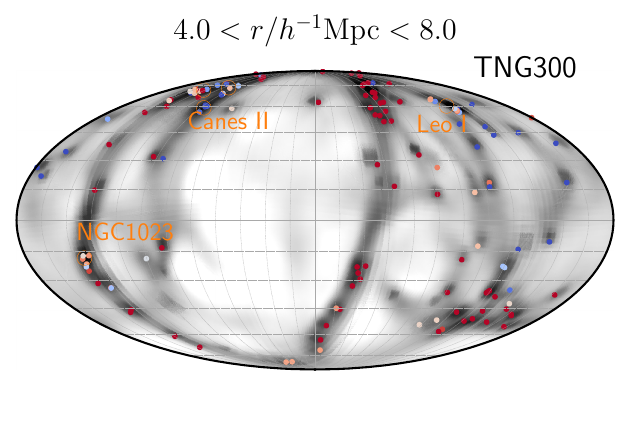} 
\hspace{0.025\textwidth} &
\hspace{0.025\textwidth}
\includegraphics[width=0.35\textwidth]{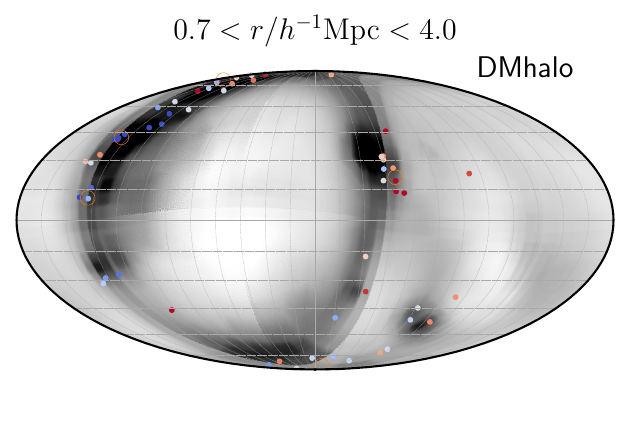}
\vspace{-15pt} \\ \cline{2-2}
\includegraphics[width=0.35\textwidth]{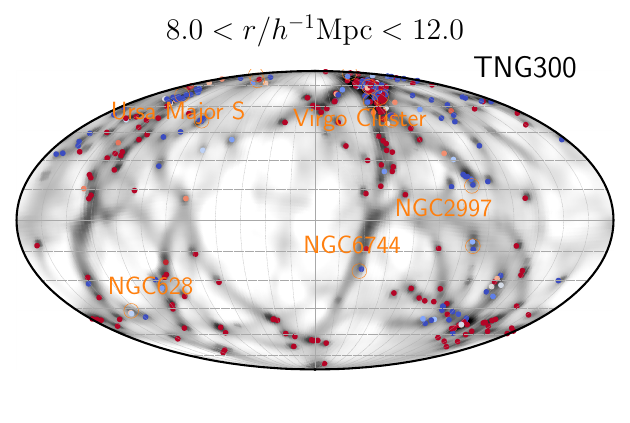}
\hspace{0.025\textwidth} &
\hspace{0.025\textwidth} 
\includegraphics[width=0.35\textwidth]{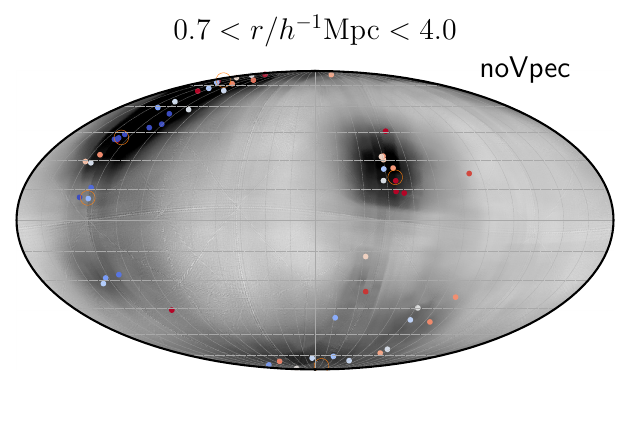}
\vspace{-15pt}  \\ 
\includegraphics[width=0.35\textwidth]{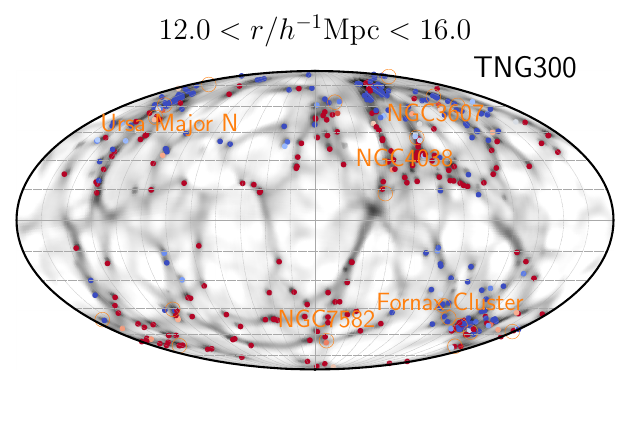}
\hspace{0.025\textwidth} &
\hspace{0.025\textwidth}
\includegraphics[width=0.35\textwidth]{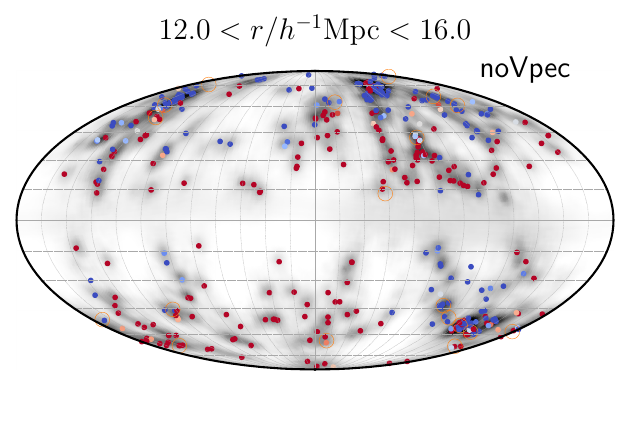}
\vspace{-15pt} \\
\includegraphics[width=0.35\textwidth]{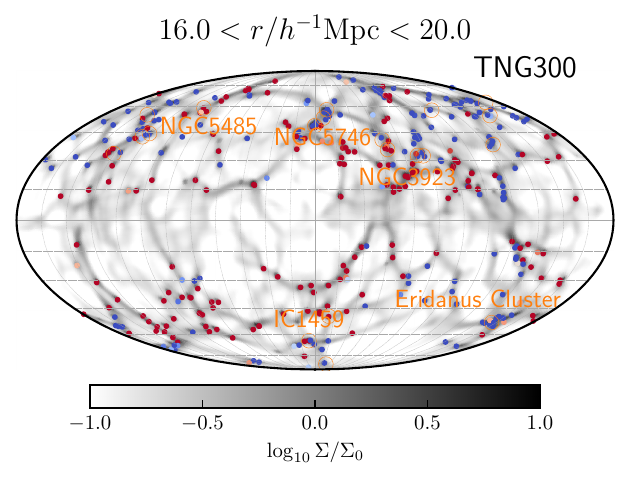}
\hspace{0.025\textwidth} &
\hspace{0.025\textwidth}
\includegraphics[width=0.35\textwidth]{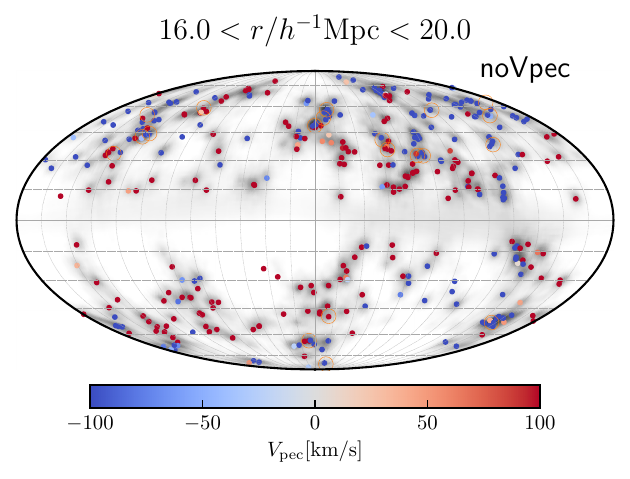}
\vspace{-5pt} \\
\end{tabular}
\caption{Two-dimensional full-sky map of the local dark-matter column density with $4\Mpch$ width\tcg{s}.
Left panels: predictions from \textsf{TNG300} training, from the nearest to the farthest radial bin.
Right panels: comparison predictions from \textsf{TNG100} training (\textsf{TNG100}), training with dark matter halos from the dark matter-only simulation (\textsf{DMhalo}), and training without using the radial peculiar velocity (\textsf{noVpec}).
Small dots: positions and peculiar velocity (color) of known local galaxies.
Large dots: galaxy groups and clusters with their names.
\label{fig:FaceMap}}
\end{figure*}

The left panels of Figure~\ref{fig:FaceMap} (labeled as \textsf{TNG300}) show the recovered local dark-matter map  
\tcg{ on the sky (gray map),
\begin{equation}
\Sigma(\vec{\theta}) \equiv \int {\rm d}r \, \rho(\vec{\theta},r) \, ,
\end{equation}
where $\vec{\theta}$, $r$, $\rho(\vec{\theta},r)$ are the two-dimensional sky coordinates, distance from the observer, and the dark-matter density at the given $(\vec{\theta},r)$, respectively.
We use the \emph{Healpix} \citep{healpy,healpix} package to reconstruct the two-dimensional sky map from the three-dimensional data cube.
We set the resolution} parameter ${\rm Nside} = 128$, which roughly corresponds to the angular resolution of $27'$\tcg{.
This figure also shows} the locations and radial peculiar velocities of galaxies that we use for the reconstruction (color-coded dots)\tcg{, as well as} the locations of some well-known galaxy groups and clusters (large dots).

\begin{figure}[hbt]
\plotone{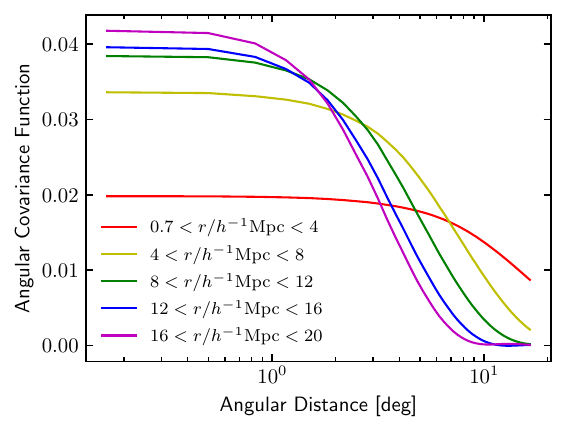}
\caption{\tcg{Angular covariance function $C(\delta \theta)$ as a function of angular distance ($\delta \theta$) in the sky.
Each angular covariance function roughly follows $C(\delta \theta) \propto \exp (-\delta \theta / \delta \theta_0)$, where $\delta \theta_0$ is a proxy of angular resolution of a degraded map each of whose pixel is statistically independent.}}
\label{fig:covmat}
\end{figure}

The map in Figure~\ref{fig:FaceMap} uses the radial distance and radial peculiar velocities reported in the \emph{Cosmicflows-3} catalog \citep{tully2016}. We have mitigated the $10-30\%$ uncertainties of distance measurement in the catalog by adopting the radial binning of $\Delta r=4\Mpch$. We further analyze the statistical uncertainties of the recovered dark-matter map by generating 1,000 realizations incorporating the uncertainties of the distance measurement \tcg{(see Section~\ref{sec:result_3d})}. From the high angular resolution map (${\rm Nside}=128$), we \tcg{define} the angular covariance function\tcg{, 
\begin{equation}
C(\delta \theta) \equiv \frac{\langle \delta \Sigma(\vec{\theta})  \, \delta \Sigma(\vec{\theta}') \rangle_{N,\vec{\theta},\vec{\theta}'}}{\Sigma_0^2} \, ,
\end{equation}
where 
$\Sigma_0 = \rho_0 \Delta r$ is the mean dark-matter column density, $\langle \ldots \rangle_{N,\vec{\theta},\vec{\theta}'}$ is the average over $N=$1,000 realizations and sky coordinates $\vec{\theta}$ and $\vec{\theta}'$ that satisfy $|\vec{\theta} - \vec{\theta}'| = \delta \theta$, and $\delta \Sigma (\vec{\theta}) \equiv \Sigma(\vec{\theta}) - \langle \Sigma(\vec{\theta} \rangle_N$.
We found that the angular covariance function follows an exponential decay over $\delta \theta$,
\begin{equation}
C(\delta \theta) \approx C_0 \exp \left( - \frac{\delta \theta}{\delta \theta_0} \right) \, ,
\end{equation}
and the values of the angular scale that shows a strong pixel-to-pixel correlation are }
$\mathbf{\delta} \theta_0  = 20.7^\circ$, $9.71^\circ$, $6.53^\circ$, $5.04^\circ$, and $4.24^\circ$, respectively, from the nearest ($r < 4\Mpch$) to the farthest ($16\Mpch<r<20\Mpch$) radial bins \tcg{(see Figure~\ref{fig:covmat}).
$\delta \theta_0$ at different radial bins correspond to the linear scales $\delta \ell = 0.26$, $0.68$, $0.92$, $1.06$, and $1.18\Mpch$, respectively.
$\delta \ell$ at the nearest radial bin well represent the spatial resolution of the three-dimensional grid ($0.3215\Mpch$). 
On the other hand, $\delta \ell$ at the farthest radial bin may mean a typical scale of the filamentary structure at given radial bin width and galaxy number density.} For the statistical analysis, we degrade the angular resolution of each map to $\mathbf{\delta} \theta_0$\tcg{---${\rm Nside} = 4$, $8$, $8$, $16$, and $16$ from the nearest to the farthest radial bins---} and assume that each pixel in the degraded map is statistically independent. 

\begin{figure*}[hbt]
\centering
\includegraphics[width=0.325\textwidth]{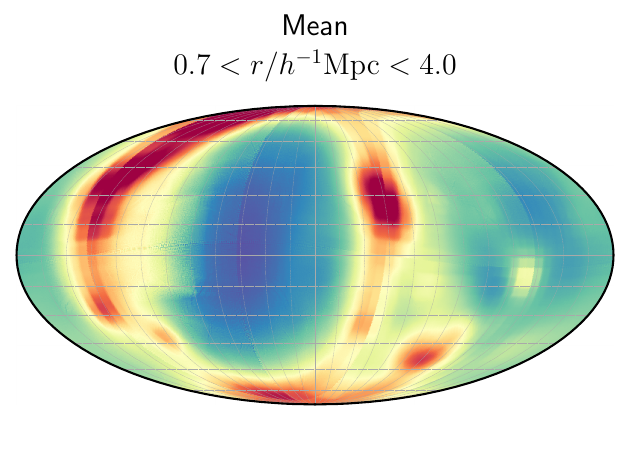}
\includegraphics[width=0.325\textwidth]{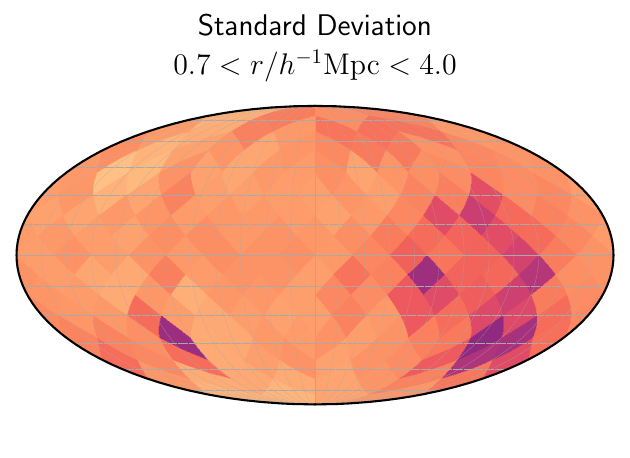}
\includegraphics[width=0.325\textwidth]{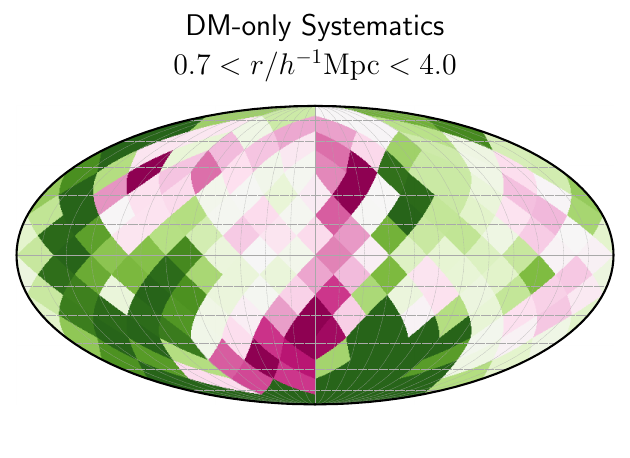}\\
\vspace{-5pt}
\includegraphics[width=0.325\textwidth]{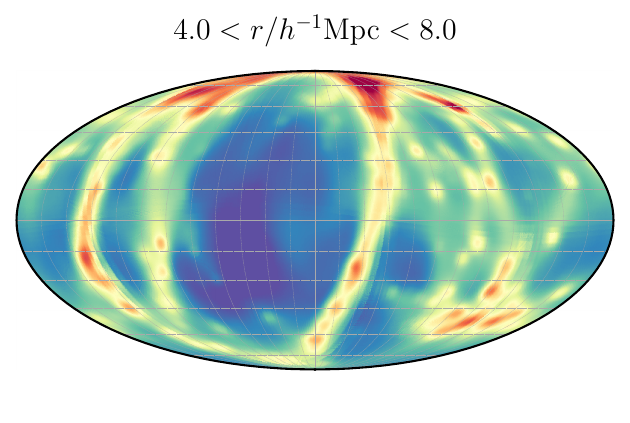}
\includegraphics[width=0.325\textwidth]{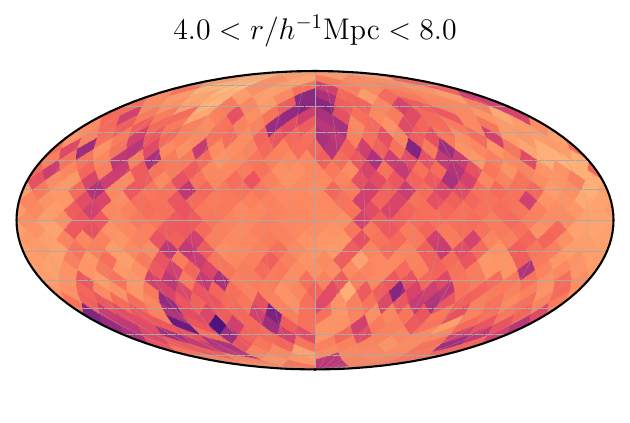}
\includegraphics[width=0.325\textwidth]{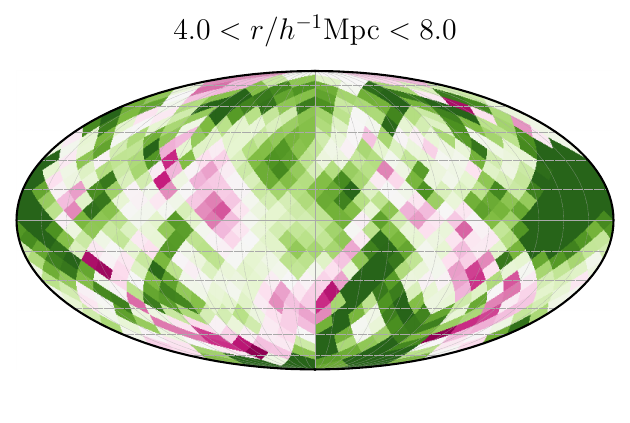}\\
\vspace{-5pt}
\includegraphics[width=0.325\textwidth]{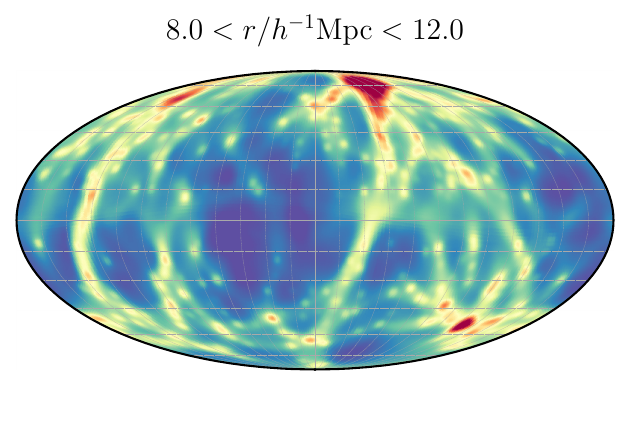}
\includegraphics[width=0.325\textwidth]{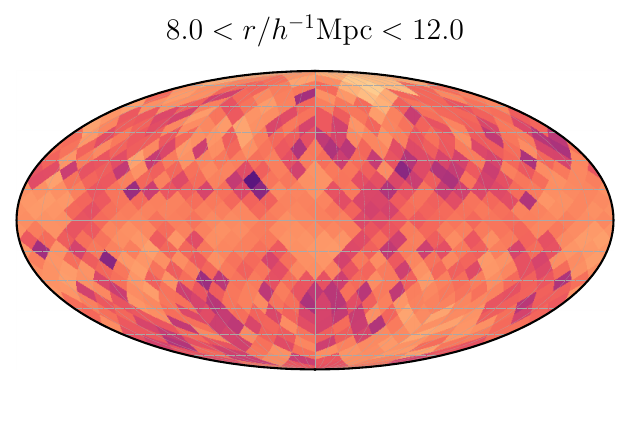}
\includegraphics[width=0.325\textwidth]{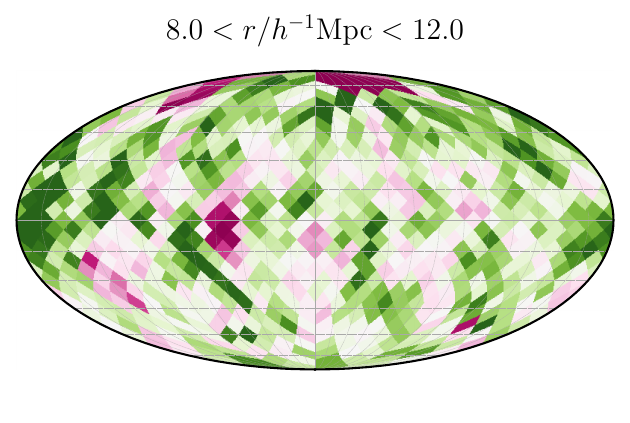}\\
\vspace{-5pt}
\includegraphics[width=0.325\textwidth]{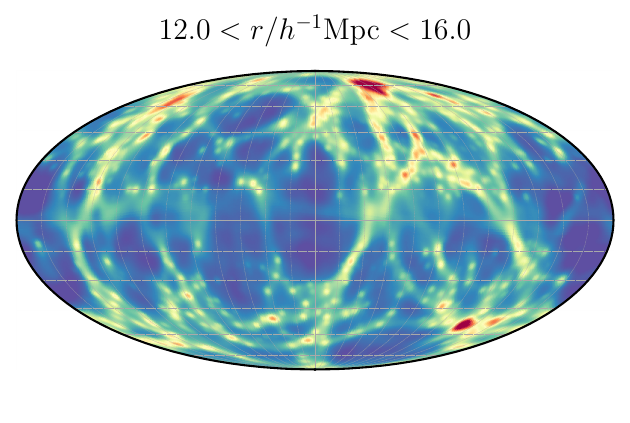}
\includegraphics[width=0.325\textwidth]{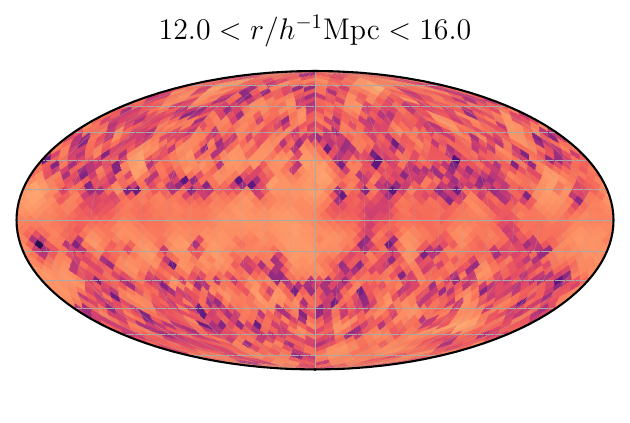}
\includegraphics[width=0.325\textwidth]{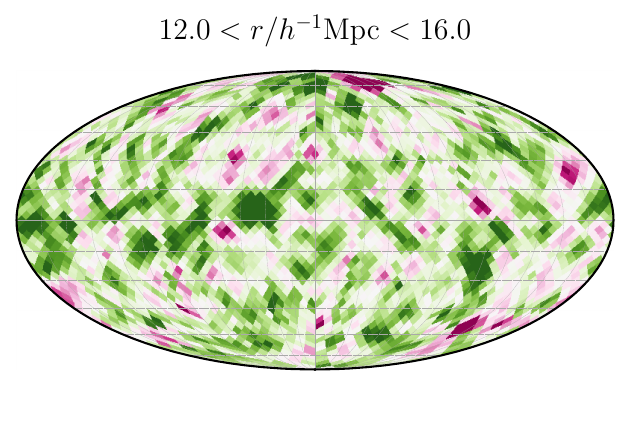}\\
\vspace{-5pt}
\includegraphics[width=0.325\textwidth]{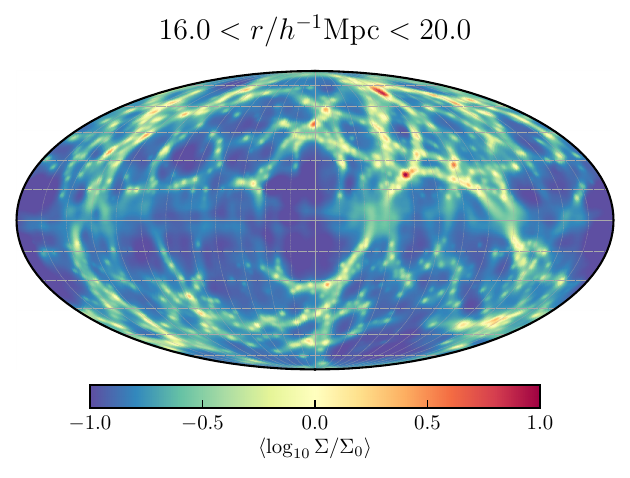}
\includegraphics[width=0.325\textwidth]{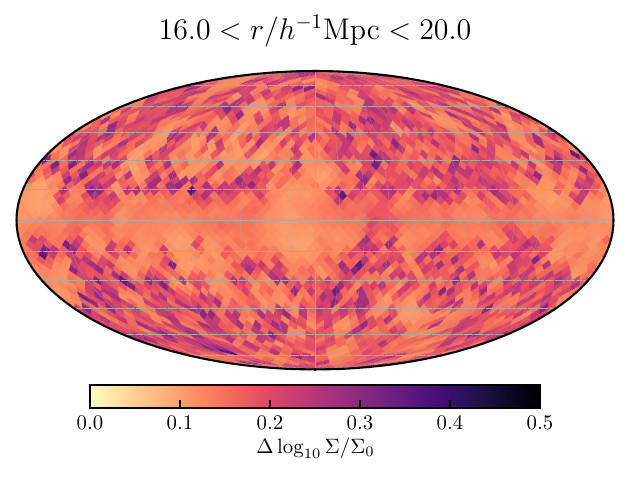}
\includegraphics[width=0.325\textwidth]{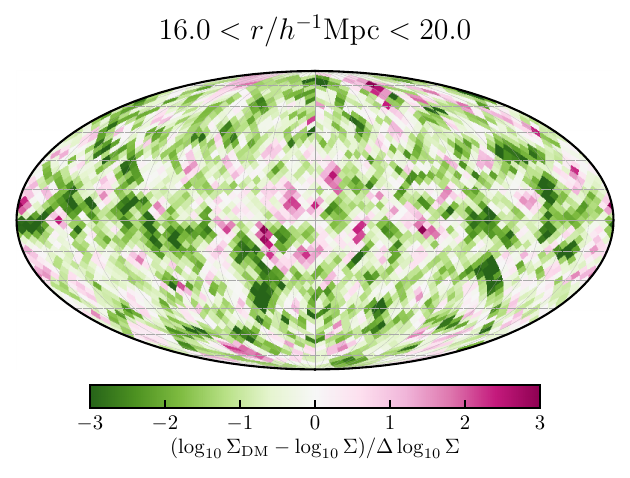}\\
\vspace{-5pt}
\caption{Same as Figure~\ref{fig:FaceMap}, but showing statistical maps. Left panels: mean of the logarithm of dark-matter column-density estimated from 1000 random realizations incorporating the uncertainties in distance estimate to the local galaxies. Middle panels: standard deviation from 1,000 random realizations (${\rm Nside}=4$, 8, 8, 16, 16 from top to bottom). Right panels: systematic bias from different simulation input for the deep-learning (\textsf{TNG300} 
\tcg{ vs. \textsf{DMhalo}}).
\label{fig:StatMap}}
\end{figure*}

Figure~\ref{fig:StatMap} shows the mean \tcg{($\langle \log_{10} \Sigma \rangle$; left panel)} and the 
\tcg{standard deviation ($\Delta \log_{10} \Sigma$; middle panel) of the logarithm of the local dark-matter column density over 1,000 realizations incorporating the uncertainties of the distance measurement.} We find that the standard deviation per pixel stays in the range of \tcg{$\Delta \log_{10} \Sigma/\Sigma_0 \simeq 0.1-0.4$, with only a mild dependence to the density contrast.
As a result, } the signal-to-noise ratio \tcg{${\rm SNR} \equiv |\langle \log_{10} \Sigma \rangle| / \Delta \log_{10} \Sigma$} scales almost linearly as the density contrast, reaching up to $\mathbf{{\rm SNR} \simeq} 10$ for the density peaks. On average, the signal-to-noise ratios for dark-matter distribution per pixel at higher Galactic latitudes ($|b| > 10^\circ$) are 4.25, 3.76, 3.94, 4.19, and 4.52, respectively, from the nearest to the farthest radial bin.

\begin{deluxetable*}{lccccc}
\tablecaption{\tcg{On-sky average (median and $1$-$\sigma$ certainty level in the parenthesis) of the systematics $\Delta_{\rm sys} \equiv |\log_{10} \Sigma - \log_{10} \Sigma_{\mathsf{TNG300},\mathrm{face}} |/\Delta \log_{10} \Sigma_{\mathsf{TNG300}}$ over high Galactic latitude $|b| > 10^\circ$ with different radial bins.
See Table~\ref{tab:models} for the definition of each comparison model except \textsf{TNG100}.}}\label{tab:systematics}
\tablehead {
\colhead{Comparison Model} &
\colhead{$0.7-4\,{\rm Mpc}/h$} &
\colhead{$4-8\,{\rm Mpc}/h$} &
\colhead{$8-12\,{\rm Mpc}/h$} &
\colhead{$12-16\,{\rm Mpc}/h$} &
\colhead{$16-20\,{\rm Mpc}/h$}
}
\startdata
\textsf{TNG100} & 2.281 ($1.837^{+1.993}_{-1.104}$) & 1.474 ($1.196^{+1.414}_{-0.842}$) & - & - & - \\
\textsf{diffH0} & 0.212 ($0.171^{+0.223}_{-0.115}$) & 0.162 ($0.133^{+0.148}_{-0.092}$) & 0.154 ($0.116^{+0.161}_{-0.083}$) & 0.152 ($0.117^{+0.153}_{-0.082}$) & 0.160 ($0.128^{+0.151}_{-0.092}$) \\
\textsf{16mag} & 1.032 ($0.949^{+0.748}_{-0.647}$) & 1.093 ($0.868^{+1.089}_{-0.611}$) & 0.862 ($0.716^{+0.729}_{-0.508}$) & 0.785 ($0.641^{+0.751}_{-0.455}$) & 0.804 ($0.631^{+0.790}_{-0.443}$) \\
\textsf{17mag} & 1.178 ($0.901^{+1.081}_{-0.572}$) & 1.105 ($0.889^{+1.026}_{-0.621}$) & 1.001 ($0.815^{+0.947}_{-0.575}$) & 0.887 ($0.726^{+0.862}_{-0.502}$) & 0.898 ($0.734^{+0.833}_{-0.506}$) \\
\textsf{noVpec} & 1.935 ($1.715^{+1.919}_{-1.359}$) & 1.105 ($0.834^{+1.120}_{-0.631}$) & 0.943 ($0.701^{+0.890}_{-0.524}$) & 0.828 ($0.672^{+0.751}_{-0.470}$) & 0.750 ($0.626^{+0.742}_{-0.440}$) \\
\textsf{stellarMass} & 1.544 ($1.256^{+1.435}_{-0.843}$) & 1.175 ($0.946^{+1.156}_{-0.684}$) & 0.925 ($0.734^{+0.909}_{-0.521}$) & 0.877 ($0.692^{+0.837}_{-0.485}$) & 0.907 ($0.713^{+0.899}_{-0.490}$) \\
\textsf{DMhalo} & 1.737 ($1.154^{+2.253}_{-0.863}$) & 1.445 ($1.127^{+1.414}_{-0.816}$) & 1.176 ($0.913^{+1.097}_{-0.610}$) & 1.057 ($0.846^{+1.029}_{-0.595}$) & 0.957 ($0.796^{+0.889}_{-0.574}$) \\
\enddata
\end{deluxetable*}

Note that, \tcg{in addition to the distance measurement uncertainty, there are} systematic uncertainties in DL mapping itself into the error budget. For example, the galaxy simulations with different resolutions or different sub-grid prescriptions can lead to different DL mapping. 
\tcg{We check such systematic effect by comparing \textsf{TNG300} with various comparison models, including those already introduced in Table~\ref{tab:models}.
To do this, we calculate the on-sky average of the systematics
\begin{equation}
\Delta_{\rm sys} \equiv \frac{|\log_{10} \Sigma - \log_{10} \Sigma_{\mathsf{TNG300}}|}{\Delta \log_{10} \Sigma_{\mathsf{TNG300}}} \, ,
\end{equation}
where $\Sigma$ and $\Sigma_{\mathsf{TNG300}}$ are the local dark-matter column densities from a given comparison model and \textsf{TNG300}, both by adopting the reported values of galaxy locations and peculiar velocities.} 
First, we check the systematic effect of the resolution by comparing the local dark-matter map estimated from \textsf{TNG100} and \textsf{TNG300}. The top-right panel of Figure~\ref{fig:FaceMap} shows the $r<4\Mpch$ bin dark-matter map driven from the \tcg{high-resolution result (}\textsf{TNG100}\tcg{)}.
\tcg{\textsf{TNG100}} systematically underestimate the density contrast by 
\tcg{$\Delta_{\rm sys} = 2.3$} on average \tcg{(see Table~\ref{tab:systematics}). 

Secondly, }to estimate the systematic effect from different sub-grid prescriptions, we have repeated the deep-learning procedure by using the dark-matter halo samples from the \tcg{dark-matter-only simulation \textsf{TNG300-1-Dark}} \tcg{by matching the galaxy/halo number density (\textsf{DMhalo})}.  
The right panels of Figure~\ref{fig:StatMap} show the difference between the two dark-matter maps in units of standard deviation at each pixel. Even with this extreme comparison \tcg{between} full hydrodynamic simulation
\tcg{ and } pure $N$-body simulation
, we find that systematic effects lead to 
\tcg{$\Delta_{\rm sys} = 1.7$, $1.4$, $1.2$, $1.1$, and $1.0$ on average} from the top (nearest) to the bottom (furthest) maps. 

We further test the systematic effect due to different Hubble parameters ($H_0=75\,{\rm km/s/Mpc}$\tcg{; \textsf{diffH0}}) and find only \tcg{$\Delta_{\rm sys} \simeq 0.15$.} Different $B$-band magnitude cuts ($M_B<-16$ \tcg{and $-17$; \textsf{16mag} and \textsf{17mag}, respectively}) \tcg{ and using total stellar mass instead of galaxy number (\textsf{stellarMass})} lead to \tcg{$\Delta_{\rm sys} \simeq 1$.} Most importantly, none of the systematic maps shows a significant correlation with the derived cosmic web structure, ensuring the robustness of the derived dark-matter distribution, or the Cosmic Web \tcg{(see the right panel of Figure~\ref{fig:StatMap})}.

The most striking feature that we have recovered in this study is the filamentary Cosmic Web that is apparent in Figures~\ref{fig:FaceMap} and \ref{fig:StatMap}. \tcg{First of all,} we find that the radial peculiar velocity information is vital to reconstructing the cosmic web, without which the same DL algorithm can not reproduce the Cosmic Web structure at all. For example, the right panels in Figure~\ref{fig:FaceMap}, indicated by \textsf{noVpec}, show the deep-learning result only using galaxy distributions. Note the absence of the filamentary structure in those maps. We note that the \textsf{noVpec} maps resemble the smoothed version of the galaxy distribution. The deep-learning algorithm with stellar-mass weighted galaxy distribution, without peculiar velocity information, leads to the similarly poor quality map. 

Another interesting feature in the map is the dark-matter distribution at lower Galactic latitudes ($|b|<10^\circ$) where we do not have any input galaxy data. To our surprise, we find that the \tcg{averaged} signal-to-noise ratios per pixel for this region are $4.18$, $4.73$, $5.31$, $5.80$, and $6.21$, respectively, from the nearest to the farthest radial bin. We, however, anticipate that the theoretical uncertainties for the DL mapping would be most substantial for this region. For example, from the aforementioned studies on systematic uncertainties, we find that\tcg{,} on average\tcg{,} lower Galactic latitudes ($|b|<10^\circ$) map suffers about \tcg{$\delta \Delta_{\rm sys} \simeq 0.5$} more systematical shifts than higher Galactic latitudes ($|b|>10^\circ$) map. This is indicated in the top two panels of Figure~\ref{fig:FaceMap} \tcg{and} the systematic shifts shown in the right panels of Figure~\ref{fig:StatMap}.

\section{Discussion}\label{sec:discuss}
In this paper, we present a novel \tcg{convolutional neural network (CNN)}
-based deep learning \tcg{(DL)} method of reconstructing the local dark-matter distribution map and discover the local Cosmic-Web structure traced by the positions and radial peculiar velocities of \emph{Cosmicflow-3} galaxies. We find that including the radial peculiar velocity field is the key to recover the dark matter distribution in the Cosmic Web. Incorporating the observational uncertainties in the galaxy distance measurements, the average detection significance of the dark-matter map exceeds $4.1$-$\sigma$ for each \emph{Healpix} pixel at higher Galactic latitudes ($|b| > 10^\circ$). The quoted statistical significance, however, does not include the uncertainties in the galaxy-to-dark-matter mapping itself. 
We have tested that the DL results stay robustly for three different simulations: \tcg{\textsf{TNG100-1} and \textsf{TNG300-1} from the \emph{Illustris-TNG} simulation and \textsf{RefL0100N1504} from the \emph{EAGLE} simulation,} but future studies must quantify the theoretical uncertainties by applying the same method to the large-scale structure simulations with different baryonic prescriptions. The comparison of the DL results between \tcg{\textsf{TNG300-1}} and $N$-body simulations, however, indicates that the filamentary Cosmic-Web structure may not suffer from the systematic effects.

The main statistical uncertainty in the galaxy data comes from the uncertainty in the distance measurement. As the observed shift in the galaxy spectra constrains the sum of the distance (Hubble flow) and the radial peculiar velocity, the uncertainty affects both the galaxy distribution and the radial peculiar velocity field. Therefore, to obtain a dark-matter map with higher significance, it is necessary to explore the ways to reduce the uncertainties of the current distance estimators such as the Tip of the Red Giant Branch, the Type Ia supernova, and the Fundamental Plane through continuous cross\tcg{-}calibration \citep{tully2016}, and to increase the number of galaxies with measured distances through systematic surveys (e.g., \emph{6dFGS} \citep{6dfgs}, \emph{James Webb Space Telescope} \citep{jwst}).

We anticipate that the reconstructed three-di\-men\-sio\-nal dark-matter map and peculiar velocity field will open an entirely new chapter of cosmological study. For example, the dark-matter map can make it possible to run the cosmological galaxy simulations with the {\it precise} initial condition of the Local Group for studying the past and future of our cosmic neighborhood. It will also allow the in-depth study of the nature of dark matter by cross-correlating the reconstructed dark matter map with the full-sky diffuse emission maps constructed from the radio-to-gamma-ray electromagnetic spectra as well as the full-sky map of gravitational wave binaries. The latter can test the models where black holes in binaries have formed out of dark matter \citep{shandera/etal:2019}.

\tcg{Finally, as we have introduced a novel CNN-based DL method to reconstruct the local Cosmic Web, the quantitative study comparing the prediction power of the DL method presented here with pre-existing methods such as BORG may be in order. Note that, however, many previous studies reconstruct the dark matter distribution on sales much larger than the size of our local Cosmic Web \citep[e.g., $\gtrsim 3-5\Mpch$ in][]{jasche2013,jasche2015}, which complicates the direct comparison between the two methods.
Nevertheless, an additional study that applies the existing methods to similar observational and simulation data to ours and compares them to our DL method would be beneficial, and we leave it for the future.}

\acknowledgements
{Authors acknowledge Christophe Pichon, Changbom Park, Sungryong Hong, Inkyu Park, Dongsu Bak, Graziano Rossi, \tcm{and} Yung-Kyun Noh for discussion. \tcg{Authors also acknowledge an anonymous referee for suggestions to improve this article.} The list of nearby galaxy groups and clusters \tcg{is} derived from \url{www.atlasoftheuniverse.com}.
Authors acknowledge the Korea Institute for Advanced Study for providing computing resources (KIAS Center for Advanced Computation Linux Cluster System).
Computational data were transferred through a high-speed network provided by the Korea Research Environment Open NETwork (KREONET).

SEH was \tcg{partly} supported by Basic Science Research Program through the National Research Foundation of Korea funded by the Ministry of Education (2018\-R1\-A6\-A1\-A06\-024\-977).
\tcg{SEH was also partly supported by the project \begin{CJK}{UTF8}{mj}우주거대구조를 이용한 암흑우주 연구\end{CJK} (``Understanding Dark Universe Using Large Scale Structure of the Universe''), funded by the Ministry of Science.}
DJ was supported at Pennsylvania State University by NSF grant (AST-1517363) and NASA ATP program (80NSSC18K1103).
JK was supported by a KIAS Individual Grant (KG039603) via the Center for Advanced Computation at Korea Institute for Advanced Study.
 }

\software{
\tcm{HEALPix \citep{healpix}, 
Healpy \citep{healpy}, 
astropy \citep{astropy2013,astropy2018}, 
NumPy \citep{numpy2011,numpy}, 
Scipy \citep{scipy2001,scipy}, 
matplotlib \citep{matplotlib}, 
pandas \citep{pandas}, 
Keras \citep{chollet2015}, 
Tensorflow backend \citep{abadi2015}}}

%\newpage

\bibliography{reference}

\end{document}